# Closed-Loop Thermo-Kinetic Valorization of Fermentation Residues into Catalytic Biochar for Enhanced Biohydrogen Production

Muhammad Shahzaib[1]

[1]Collaborative Innovation Center of Biomass Energy, Henan Agricultural University, Zhengzhou 450002

Corresponding author: Muhammad Shahzaib

engr.shahzaib04899@yahoo.com

**Abstract**

The current research critically evaluates the closed-loop biorefinery approach, in which fermentation residues (FRs) from photo-fermentative biohydrogen production (PFHP) are further upcycled thermochemically into functional biochar catalysts, thereby enhancing the efficiency of the original PFHP approach. Four different types of FRs were obtained after processing hydrothermally and ethylene glycol-pretreated corn stover and pyrolyzing it at 700 °C. Kinetic analysis by the Coast-Redfern model suggested a D3/D4 diffusion mechanism, while the KAS, FWO, and Starink models evaluated activation energies in the range of 157-278 kJ/mol for the reaction results. The thermodynamic parameters analysis confirmed the choice of feedstock on the spontaneity of reaction and entropy balance in the reaction systems. Ultimate and proximate analysis confirmed the higher dependency of pyrolysis product distribution on FRs composition. The pyrolysis process restored the porosity that was lost due to the previous fermentation process, thus yielding biochar with specific characteristics generated either as microporous BC3 with the highest specific surface area of 185.14 m²/g from oxygenated feedstock, or mesoporous biochar BC4 from graphitized FRs with a specific surface area of 76.58 m²/g. When added to the PFHP process, BC3 was seen to maximize the cumulative hydrogen production (570 mL) due to its higher pH buffering capacity. The addition of BC4 allowed for the highest yield of hydrogen production rate of 14.91 mL/h as a result of increased electron shuttling mechanisms. The entire approach of FRs valorization optimally indicated a method of process enhancement with minimal waste outputs, within a context of a renewable energy framework and Circular Bioeconomy.

# 1 Introduction

The escalating global energy crisis and concomitant environmental degradation necessitate an urgent transition towards sustainable, carbon-neutral energy systems [1]. Within this context, biohydrogen production via photo-fermentation of lignocellulosic biomass (PFHP) represents a highly promising clean energy carrier, characterized by its high energy density and zero carbon emissions upon combustion [2-4]. However, the economic viability and environmental sustainability of large-scale PFHP implementation are significantly constrained by the substantial quantities of fermentation residues (FRs) generated as process waste [5]. Conventional approaches to FR management, such as utilizing feedstocks for lipid accumulation [6, 7], seed growth media [8], or mushroom substrates [9], present inherent limitations. For sure, inefficient management can lead to environmental pollution and impose additional energy burdens. Furthermore, regulatory restrictions on the direct composting of fermentation residues, sludges, and analogous waste streams have been enacted in several authorities [10]. Present disposal methods, such as landfilling and incineration, contribute to soil contamination, greenhouse gas emissions, and resource wastage, thereby undermining the fundamental sustainability benefits for biohydrogen production [5]. Conversely, the effective valorization of FRs holds transformative potential for sustainable biorefineries by enabling resource circularity, waste minimization, and reducing feedstock costs [10, 11].

Recent investigations into cyclic PFHP systems underscore efforts to maximize feedstock utilization. Shahzaib et al. demonstrated that FRs derived from corn stover PFHP retained 36.5% cellulose after two cycles; however, a concomitant surge in lignin content to 22.4% impeded further biochemical reuse [5]. Similarly, Jiang et al. observed a 40% decline in biohydrogen yield during the third PFHP cycle, which was attributed to the accumulation of lignin [12]. These findings collectively indicate that after 2-3 cycles, reutilization of these lignin-rich FRs is not possible through biochemical processes. However, thermochemical conversion pathways, notably pyrolysis, offer a viable alternative, transforming FRs into bio-oil, syngas, and biochar [13, 14]. Pyrolysis discusses significant advantages, including substantial volume reduction through lignin degradation, complete pathogen elimination, and facilitation of heavy metal precipitation and complexation [15, 16]. However, during this process, macromolecular organic matter (lignin) within FRs is predominantly converted into liquid bio-oils [17]. The resulting gaseous products (hydrogen and methane) serve as fuel gas, while bio-oils can be upgraded into liquid biofuels or chemicals; the solid residue (biochar) constitutes a valuable co-product [15, 16].

Biochar derived from FRs exhibits versatile environmental applications, including soil enhancement [18], wastewater remediation [19], and the removal of antibiotics [20]. Notably, it has also demonstrated efficacy as a catalyst in biohydrogen production processes [21]. For instance, Yang et al. reported that activated biochar from corncob xylose residue possesses an ultra-high surface area (3043 m²/g) and exceptional adsorption capacity (1429 mg/g) for antibiotics [20]. Furthermore, stillage residue-derived biochar achieved a maximum silver ion adsorption capacity of 90.06 mg/g from aqueous solutions [22]. Biochar from FRs has been successfully deployed as a soil remediation agent [18] and a heavy metal adsorbent [23]. Most significantly, FRs-based biochar's exceptional physicochemical properties, including high surface area, contaminant immobilization capacity, promotion of microbial growth, nutrient availability, buffering capability, and enhanced electron transfer efficiency, render it an outstanding catalyst candidate for biohydrogen production. These attributes directly address critical bottlenecks in improving biohydrogen production [24-27]. Therefore, the pyrolytic conversion of FRs into biochar presents an effective strategy for mitigating environmental burdens and enabling the recycling of biomass energy. A fundamental challenge persists: the heterogeneous nature of FRs, dictated by feedstock origin, pretreatment methods, and microbial consortia, causes complex variations in conversion kinetics, thermodynamics, and product distribution that remain inadequately characterized. Most studies focus on the pyrolysis of virgin biomass and overlook the unique physicochemical properties of FRs, which have different lignocellulosic structures, mineral content, and resistant organics due to prior processing [13, 14, 28, 29]. This gap hinders the understanding of the thermo-kinetic mechanisms and thermodynamic factors controlling FR pyrolysis, limiting the ability to design efficient, customized conversion processes that maximize resource recovery and reduce environmental impact.

This study addresses the research gap by investigating thermochemical valorization of fermentation residues (FRs) from photo-fermentative biohydrogen production (PFHP) of pretreated corn stover. It specifically examines how pretreatment-induced changes in FR composition and structure influence pyrolysis kinetics, thermodynamics, and product yields, and explores optimizing the resulting biochar as a catalyst to enhance the initial PFHP process. The novelty lies in establishing a closed-loop, zero-waste biorefinery model where FRs become valuable feedstocks for producing clean fuels (bio-oil, syngas) and engineered biochar (BC), which serves as a catalyst that significantly boosts PFHP efficiency. This holistic approach moves beyond linear models by fostering synergistic interactions between biochemical and thermochemical systems. These experimental findings provide new mechanistic insights into FR pyrolysis through multi-model kinetic analysis (Kissinger-Akahira-Sunose, Ozawa-Flynn-Wall, Starink, Coats-Redfern) and thermodynamic profiling (ΔH, ΔG, ΔS), demonstrating how pretreatment severity (hydrothermal vs. ethylene glycol) selectively modifies activation energy, reaction spontaneity, entropy, and diffusion-limited decomposition pathways. It examined how pyrolysis can overcome pore blockages induced by fermentation, transforming damaged FRs into biologically functional biochar. It also examined how precursor type determines biochar properties, such as microporosity, graphitic order, surface chemistry, and mineral composition, which influence their catalytic ability for PFHP enhancement.

FRs are characterized for assessing composition, porosity (BET/BJH), and thermal degradation (TGA/DTG). Pyrolysis kinetics are used to determine activation energies ($E_a$) and pre-exponential factors (A), with thermodynamic parameters clarifying energy needs and reaction feasibility. BCs were extensively characterized (SEM, XRD, FTIR, BET) to link precursor properties with structural changes. These BCs were tested in PFHP reactors at optimized doses to evaluate effects on hydrogen production kinetics, total yield, pH stability, oxidation-reduction potential (ORP), and metabolite profiles. This comprehensive process, encompassing residue analysis and modeling, biochar production, and functional testing, provides a robust scientific foundation for the proposed circular economy. By converting waste into valuable process inputs, this integrated system exemplifies a closed-loop approach to PFHP, enabling zero-waste bioenergy with enhanced economic and environmental performance.

## 2 Materials and Methods

### 2.1 Feedstock preparation and nomenclature

Corn stover-based FRs from our previous study on photo-fermentation biohydrogen production were used in this study as the initial raw feedstock for evaluating thermochemical conversion [30] The physicochemical properties of FRs were altered due to the cyclic PFHP [5], followed by subsequent hydrothermal and ethylene glycol pretreatments of the initial corn stover based FRs [30]. A total of four samples, based on different severities and types of pretreatment agents, were selected (S1, S2, S3, and S4). HTP and EG represent the types of pretreatment agents, specifically hydrothermal and ethylene glycol, respectively, while 1 and 2 indicate the lower and higher severity index (180 ºC, 30 min; 180 ºC, 60 min) of pretreatment. For suitability, these samples are renamed as S1, S2, S3, and S4 instead of HTP1, HTP2, EG1and EG2, respectively. After the thermochemical conversion of these samples, biochar was obtained and named BC1, BC2, BC3, and BC4 accordingly. The type of pretreatment, the severity index, and the elemental, compositional, and structural properties of samples from the previous study are given in Table 1.

**Table 1:** Pretreatment induced changes in FRs properties examined in previous study [30]

| Sample | HTP1 | HTP2 | EG1 | EG2 |
|---|---|---|---|---|
| Severity Index | 3.83 | 4.13 | 3.83 | 4.13 |
| **Elemental Mapping** | | | | |
| N (%) | 0.4 | 0.47 | 0.52 | 0.57 |
| C (%) | 34.35 | 36.39 | 36.05 | 35.34 |
| H (%) | 3.24 | 3.15 | 3.03 | 3.39 |
| S (%) | 3.99 | 4.49 | 4.21 | 3.53 |
| C/N | 85.30 | 77.62 | 68.86 | 62.29 |
| C/H | 10.62 | 11.56 | 11.89 | 10.41 |
| **Components Present** | | | | |
| Cellulose (%) | 38.9 ± 0.7 | 40.2 ± 0.8 | 43.5 ± 0.6 | 45.5 ± 0.6 |
| Hemicellulose (%) | 12.1 ± 0.2 | 9.4 ± 0.2 | 23.1 ± 0.8 | 24.4 ± 0.9 |
| Lignin (%) | 28.4 ± 0.5 | 24.7 ± 0.4 | 17.9 ± 0.6 | 12.6 ± 0.2 |
| **Structural Characteristics** | | | | |
| $BET_{SSA}$ (m$^2$/g) | 0.395 | 0.663 | 0.265 | 0.735 |
| $BJH_{TPV}$ (cm$^3$/g) | 0.00121 | 0.00552 | 00.00127 | 0.00251 |
| $BJH_{APS}$ (nm) | 31.352 | 24.057 | 109.82 | 15.234 |

## 2.2 Theoretical Framework for Kinetic Analysis

The pyrolysis process of biomass can be represented by a first-order rate equation, as shown in Eq. (1).

$$\text{Biomass (solid)} \xrightarrow{k} \text{Volatiles + Char (Solid)} \qquad (1)$$

Here, the volatiles comprise the total tar and gas yield. The rate constant, *k*, defined by the Arrhenius equation [31], is given in Eq. (2).

$$k = A\exp\left(-\frac{E_\alpha}{RT}\right) \qquad (2)$$

In this equation, (A) is the pre-exponential factor (min$^{-1}$), (Ea) is the activation energy (kJ mol$^{-1}$), (R) is the universal gas constant (8.314 J mol$^{-1}$ K$^{-1}$), and (T) is the absolute temperature (K). The conversion rate is expressed by Eq. (3) [32].

$$\frac{d\alpha}{dt} = k(T)f(\alpha) \tag{3}$$

Here, (α) denotes the degree of conversion, (t) is time, and (f(α)) is the reaction model function. The conversion degree (α), calculated as the normalized mass loss [33], is defined in Eq. (4).

$$(\alpha) = \frac{m_o - m_t}{m_o - m_r} \tag{4}$$

Where ($m_o$) is the initial mass, ($m_t$) is the mass at time (t), and ($m_f$) is the final mass after pyrolysis. Combining Eqs. (3) and (4) yield the fundamental expression, Eq. (5), used to determine kinetic parameters from Thermogravimetric Analysis (TGA) data.

$$\frac{d\alpha}{dt} = A \exp\left(-\frac{E_\alpha}{RT}\right) f(\alpha) \tag{5}$$

For non-isothermal TGA experiments, where temperature increases linearly with time at a constant heating rate (β), Eq. (6) applies.

$$\beta = \frac{dT}{dt} = \frac{dT}{d\alpha} \times \frac{d\alpha}{dt} \tag{6}$$

From Eqs. (5) and (6), the relationship can be derived as shown in Eq. (7).

$$\frac{d\alpha}{dT} = \frac{A}{\beta} \exp\left(-\frac{E_\alpha}{RT}\right) f(\alpha) \tag{7}$$

Integration and rearrangement of Eq. (7) lead to Eq. (8) [34-36].

$$g(\alpha) = \int_0^\alpha \frac{d\alpha}{f(\alpha)} = \frac{A}{\beta} \int_{T_0}^T \exp\left(-\frac{E_\alpha}{RT}\right) dT = \frac{AE_\alpha}{\beta R} p(x) \tag{8}$$

Here, (g(α)) is the integral form of the conversion function, and (x = Ea = RT). The temperature integral (p(x)) lacks an exact analytical solution and must be approximated or solved numerically. Iso-conversional methods are primarily distinguished by the specific approximation employed for the (p(x)) function [37]. Based on Eq. (8), the kinetic parameters for pyrolysis (Ea & A) were determined from TGA data using model-free methods. This study employed the Kissinger-Akahira-Sunose (KAS), Ozawa-Flynn-Wall (OFW), and Starink model-free methods, alongside the model-based Coats-Redfern method [13, 38], which are briefly outlined below.

### 2.2.1 Kissinger–Akahira–Sunose (KAS) method

This method calculates the activation energy using the approximation $p(x) = x^2 e^{-x}$ and substituting this into Eq. (8) and rearranging yields Eq. (9) [39, 40].

$$\ln\left(\frac{\beta}{T^2}\right) = \ln\left(\frac{AR}{E_\alpha g(\alpha)}\right) - \frac{E_\alpha}{RT} \qquad (9)$$

Plotting $\ln(\beta/T^2)$ against $(1/T)$ for fixed conversion values ($\alpha$) (from 0.1 to 0.9) at different $\beta$ yields straight lines. The activation energy (Ea) for each conversion degree is calculated from the slope of these lines (-Ea/R).

### 2.2.2 Ozawa–Flynn–Wall (OFW) method

This technique is based on Doyle's approximation ($\log[p(x)] = 2.315 + 0.457x$). Substitution into Eq. (8) gives Eq. (10), which is used to compute (Ea) [40, 41].

$$\ln\beta = \ln\left(\frac{AE_\alpha}{Rg(\alpha)}\right) - 5.331 - 1.052\frac{E_\alpha}{RT} \qquad (10)$$

Plotting $(\ln/\beta)$ versus $(1/T)$ for specific conversion values yields straight lines, whose slopes (-1.052Ea/R) are used to determine the activation energy.

### 2.2.3 Starink method

The Starink method synthesizes the OFW and KAS approaches into a more accurate approximation, as shown in Eq. (11) [13, 42].

$$\ln\left(\frac{\beta}{T^{1.92}}\right) = C - 1.0008\frac{E_\alpha}{RT} \qquad (11)$$

The activation energy is derived from the slope of the plot of $(\ln(\beta/T^{1.92}))$ against $(1/T)$, which is (-1.0008Ea/R).

### 2.2.4 Coats-Redfern method

The Coats-Redfern method [43] was used to determine the pre-exponential factor (A) and identify the dominant reaction mechanism during FRs pyrolysis [44]. While the OFW, KAS, and Starink methods estimate (Ea), the Coats-Redfern method uses these (Ea) values to calculate (A) [45]. For an assumed nth order reaction model, the rate equation is given by Eq. (12).

$$\frac{d\alpha}{dT} = \frac{A}{\beta} exp\left(-\frac{E_\alpha}{RT}\right)(1-\alpha)^n \tag{12}$$

Rearranging Eq. (12) gives the expression of Eq. (13).

$$\frac{d\alpha}{(1-\alpha)^n} = \frac{A}{\beta} exp\left(-\frac{E_\alpha}{RT}\right) dT \tag{13}$$

Integration of Eq. (13) gives:

$$\frac{1-(1-\alpha)^{1-n}}{1-n} = \frac{A}{\beta} \int_{T_0}^{T} exp\left(-\frac{E_\alpha}{RT}\right) dT \tag{14}$$

The temperature integral in Eq. (14) is approximated using a Taylor series expansion, with higher-order terms ignored [41], resulting in the final form presented in Eq. (15).

$$\frac{d\alpha}{dT} = \ln[g(\alpha)] - \frac{E_\alpha}{RT} + \ln\left(\frac{AR}{\beta E_\alpha}\right) \tag{15}$$

Where 
$$g(\alpha) = -\ln\left[\frac{(1-\alpha)}{T^2}\right] \quad \text{for } n = 1 \tag{16}$$

$$g(\alpha) = \frac{1-(1-\alpha)^{1-n}}{(1-n)T^2} \quad \text{for } n \neq 1 \tag{17}$$

The activation energy (Ea) and pre-exponential factor (A) are calculated from the slope and intercept, respectively, of the linear plot of (ln [g(α)] against 1/T) for the appropriate reaction order (n) using Eq. (16, 17) [46].

Various solid-state reaction mechanism models (g(α)), summarized in Table 2, were tested. These include the Homogeneous Model (HM, O1), the Shrinking Core Model (SCM, R3), random nucleation models (A2, A3), and diffusion-controlled models (D1, D2, D3, D4). The HM (O1) assumes a chemical reaction controls the process uniformly throughout the particle [47]. In contrast, diffusion models (D1, D2, D3, D4) postulate that the rate is governed by the transport of volatiles through the solid. The SCM (R3) describes a reaction interface that moves inward as the unreacted core shrinks [44].

**Table 2:** Pyrolysis Reaction Mechanism Models

| Model | Symbol | | Mathematical Expression |
|---|---|---|---|
| **Homogeneous Model** | | | |
| First-order | O1 | f(α) | $(1-\alpha)$ |
| | | g(α) | $-\ln(1-\alpha)$ |
| **Shrinking Core Model** | | | |
| Three dimensions (Contracting Sphere) | R3 | f(α) | $3(1-\alpha)^{2/3}$ |
| | | g(α) | $1-(1-\alpha)^{1/3}$ |
| **Diffusion Models** | | | |
| One-dimensional diffusion | D1 | f(α) | $\dfrac{1}{2\alpha}$ |
| | | g(α) | $\alpha^2$ |
| Two-dimensional diffusion, cylindrical symmetry (Valensi model) | D2 | f(α) | $[-\ln(1-\alpha)]^{-1}$ |
| | | g(α) | $\alpha + (1-\alpha)\ln(1-\alpha)$ |
| Three-dimensional diffusion, spherical symmetry (Jander model) | D3 | f(α) | $\dfrac{3}{2}(1-\alpha)^{2/3}[1-(1-\alpha)^{1/3}]^{-1}$ |
| | | g(α) | $[1-(1-\alpha)^{1/3}]^2$ |
| Three-dimensional diffusion, cylindrical symmetry | D4 | f(α) | $\dfrac{3}{2}[(1-\alpha)^{-1/3}-1]^{-1}$ |

| (Ginstling model) | | g(α) | $1 - \dfrac{2\alpha}{3} - (1-\alpha)^{2/3}$ |
| --- | --- | --- | --- |
| **Random Nucleation and Growth Models** | | | |
| Avrami-Erofeev | A2 | f(α) | $2(1-\alpha)[-\ln(1-\alpha)]^{1/2}$ |
| | | g(α) | $[-\ln(1-\alpha)]^{1/2}$ |
| Avrami-Erofeev | A3 | f(α) | $3(1-\alpha)[-\ln(1-\alpha)]^{2/3}$ |
| | | g(α) | $[-\ln(1-\alpha)]^{1/3}$ |

## 2.3 Thermodynamic Parameter Assessment

The thermodynamic parameters for the pyrolysis process were evaluated using the following equations [48, 49]. The pre-exponential factor (A) was calculated using Eq. (18), where Ea is the activation energy, β is the heating rate, R is the universal gas constant, and Tp is the peak temperature from the DTG curve. The enthalpy changes (ΔH), Gibbs free energy (ΔG), and entropy (ΔS) were determined using Eqs. (19), (20), and (21), respectively. In these equations, ($T_\alpha$) is the temperature at a specific conversion, ($K_B$) is the Boltzmann constant (1.381×10$^{-23}$ J/K), (h) is Planck's constant (6.626×10$^{-34}$ J·s), and ($A_0$) is the intercept of the line.

$$A = \frac{E_a \times \beta \times \exp\left(\frac{E_a}{T_p \times R}\right)}{T_p^2 \times R} \quad (18)$$

$$\Delta H = E - R \times T_\alpha \quad (19)$$

$$\Delta G = R \times T_p \times \ln\left(\frac{K_B \times T_p}{A_0 \times h}\right) + E \quad (20)$$

$$\Delta S = \frac{(\Delta H - \Delta G)}{T_p} \quad (21)$$

## 2.4 Pyrolysis characteristics and experimental procedure

The pyrolysis behavior of samples (S1, S2, S3, S4) was analyzed using a thermogravimetric analyzer (TGA/DSC, METTLER) under a high-purity nitrogen (N$_2$,

≥99.99%) atmosphere. Approximately 10 mg of each sample was placed in an open alumina crucible and heated from 30°C to 700°C at constant heating rates of 10, 20, and 30 °C/min, with a nitrogen flow rate of 50 mL/min. Data collection and processing followed the recommendations of the ICTAC Kinetics Committee [13]. The Comprehensive Pyrolysis Index (CPI), defined in Eq. (22), was used to evaluate and compare the pyrolysis performance of the samples [50].

$$CPI = \frac{DTG_{max} \times DTG_{mean} \times M_f}{T_i T_p \Delta T_{1/2}} \qquad (22)$$

Here, ($T_i$) is the initial decomposition temperature, ($M_f$) is the final residual mass, ($T_p$) is the peak temperature, ($DTG_{max}$) is the maximum mass loss rate, ($DTG_{mean}$) is the average mass loss rate, and ($\Delta T_{1/2}$) is the temperature range at which the reaction rate is half of its maximum value. The unit of CPI is $\%^2$ min$^{-2}$ °C$^{-3}$.

## 2.5 Biochar Production

Biochar was produced at lab-scale in a horizontal tube furnace (Lantian, Hangzhou). For each run, 30 g of dried FRs (S1, S2, S3, S4) were placed in a quartz boat positioned at the center of the furnace. A constant flow of $N_2$ (150 mL min$^{-1}$) was maintained as the carrier gas. The furnace started heating from room temperature to 700°C at a rate of 10°C min$^{-1}$ and sustained at the target temperature for 2 hours. Condensable vapors (bio-oil) and non-condensable gases (syngas) were collected using a water-cooling system attached to the furnace outlet. After pyrolysis, the system was cooled to room temperature under a continuous flow of $N_2$. Furthermore, the resulting biochar was collected, washed with deionized water to remove impurities, dried, ground, and sieved to a particle size of <75 μm for subsequent analysis and application.

## 2.6 Characterization of Fermentation Residues and Biochar

The morphological structural properties of FRs-based biochar were assessed by a number of technical methods. X-ray Diffraction (XRD) profiles were taken on a Bruker D8 Advance diffractometer with 40 kV/30 mA CuKα (λ= 1.5406 Å) [51]. Scanning Electron Microscopy (SEM) of surface structures was carried out by using a ZEISS Gemini 300 electron microscope to evaluate surface microstructures [52, 53]. Specific surface area, pore volume, and pore size distributions were determined via $N_2$ adsorption/desorption isotherms taken on a MicrotracBel BELSORP-mini II analyzer. The adsorption isotherm technique underlying the Brunauer-Emmett-Teller (BET) analysis was utilized to obtain the specific surface area. At the same time, pore size distribution obtained from adsorption isotherm data using Barrett-Joyner-Halenda (BJH) profiles [54]. Characterization of surface functional groups was undertaken by Fourier-Transform Infrared (FTIR) spectroscopy and was determined by methods used in previous studies [52, 53].

Proximate quantitative analysis of the all FRs was undertaken in order to obtain the following parameters: moisture content (MC), volatile matter content (VM), ash content (AC), and fixed carbon (FC). The moisture content was determined by drying at 105 °C to constant weight (ASTM E-871). VM was determined by heating a dried sample in a covered crucible at 950°C for 7 minutes (ASTM E-872). AC was found by combusting the moisture-free sample in an open crucible at 550-600°C for 4 hours (ASTM E-1755). FC was calculated by difference: FC = 100% - (MC% + VM% + AC%) [55-57]. Ultimate analysis was performed using an elemental analyzer (Elementar Analysensysteme GmbH, Germany) to determine the weight percentages of carbon (C), hydrogen (H), nitrogen (N), and sulfur (S) [5]. Oxygen (O) content was calculated by difference: O% = 100% - (C% + H% + N% + S% + Ash%). The higher heating value (HHV) and lower heating value (LHV) were estimated empirically from the ultimate analysis data using the following Eqs. (23) and (24), respectively [58-60].

$$HHV = 0.341C + 1.323H + 0.0685 - 0.0153AC - 0.1194 (O + N) \qquad (23)$$

Here, C, H, O, N, and AC represent the sample's carbon, hydrogen, oxygen, nitrogen, and ash content, respectively, in mass%.

$$LHV\ (MJ\ kg-1) = HHV - hg\ (9H/100 + MC/100) \qquad (24)$$

LHV, HHV, H, MC, and hg are the lower heating value, higher heating value, hydrogen%, moisture content%, and latent heat of steam (hg=2.26 MJ kg$^{-1}$), respectively.

## 2.7 Inoculum preparation and PFHP assay for biochar evaluation

A pure strain of purple non-sulfur bacteria (PNSB), obtained from the Nano-Photocatalytic Materials Laboratory at Henan Agricultural University, was used as the inoculum. The inoculum was cultivated using the methodology described in our previous work [5, 61, 62]. Biohydrogen production experiments were conducted in 250 mL batch reactors containing 200 mL of biohydrogen production medium, 2.5 mL of cellulase, 50 mL of PNSB inoculum (initial optical density of 0.875 g/L), 6 g of crushed corn stover, and varying doses of biochar (0, 20, 40, 60, 80, 100 mg/L). The reactors were maintained at 30 ± 2°C under continuous illumination (190 W/m²). [5, 61, 62]. Standard protocols from earlier studies were followed for corn stover preparation, biohydrogen production media formulation, gas collection and measurement, and analysis of fermentation broth parameters (pH, oxidation-reduction potential (ORP), reducing sugar (RS), and metabolic byproducts (VFAs)) [5, 61, 62].

## 3 Results and Discussion

### 3.1 FRs Characteristics

#### 3.1.1 Proximate and elemental analysis

The composition of S1, S2, S3, and S4 is shown in Table 3 with elemental analysis (Carbon%, Hydrogen%, Nitrogen%, Sulfur%, and Oxygen% (CHNSO)) and proximate analysis parameters. Among all samples, S2 has the lowest moisture content (MC) of 1.62%, requiring less energy for water evaporation. In contrast, S4 may experience a decrease in thermal efficiency due to its higher 5.46% MC.

Also, S2 has the highest volatile content (VC) of 65.85%, improving the overall yield of bioenergy products (gases, bio-oil). However, S4 has a lower 60.09 % VC, allowing it to be better suited to produce a higher biochar yield [13, 63]. In the case of ash content (AC), S1 has the highest AC, while S2 produced a low AC, and S3 has even less AC, showing that both are better suited for higher pyrolytic performance than S1. The S2 and S3 exhibit the highest fixed carbon (FC), showing a stable production of biochar with these FRs. However, the higher heating values (HHV) of S1 and S4 are of comparative and higher values than S2, which further indicate a higher energy value of their carbon-rich nature. All samples show moderate carbon contents and similar hydrogen contents; however, S4 shows the least hydrogen to carbon ratio, thus being the most aromatic stable biochar for future use. S3 shows the highest oxygen content, which is an indication of highly oxygenated bio-oil, possibly of poor stability with high acidity. Respecting the air toxicity of pyrolysis, S4 shows the lowest values in sulphur and nitrogen contents, showing its greater suitability as a feedstock in pyrolysis due to the decrease in NOx and Sox pollutants. According to the values of the carbon-to-nitrogen ratio, due to the highest value of sample S1 shows a lower decomposition rate, thus favoring its biochar production. Further, S3's high (O+N)/C value indicates the possible production of highly polarized biochar of increased capacity in adsorption. After the analysis of all parameters present in Table 3 for pyrolysis, results suggested that the S1 will not favor further except when either catalytically improved upon or pretreatments are applied for its higher AC and S% reduction. The sample S2 is followed as the most qualifying for bio-oil owing to its higher VC with lower AC and MC present, while S3 is preferred for gas production because of its highest O%, which might lead to lower-quality bio-oil. Notably, S4 shows a well-balanced profile, indicating its dual ability to produce both biochar and energy products. [13, 63, 64]. To further investigate the hypothesis mentioned above, all samples were thoroughly examined using different techniques and methods.

Table 3: The heating values, ultimate analysis and proximate analysis of FRs

| Parameter | S1 | S2 | S3 | S4 |
|---|---|---|---|---|
| **Proximate Analysis** | | | | |
| MC | 3.86 | 1.62 | 5.24 | 5.46 |
| VC | 61.69 | 65.85 | 62.94 | 60.09 |
| AC | 27.57 | 16.71 | 16.71 | 22.95 |
| FC | 6.88 | 15.82 | 15.11 | 11.5 |
| HHV | 9.17 | 8.9 | 8.14 | 8.96 |
| LHV | 8.42 | 8.22 | 7.38 | 8.19 |
| **Elemental Analysis** | | | | |
| N (%) | 0.48 | 0.62 | 0.72 | 0.54 |
| C (%) | 34.99 | 34.29 | 32.94 | 34.96 |
| H (%) | 3.23 | 3.16 | 3.14 | 3.17 |
| S (%) | 5.86 | 5.8 | 4.82 | 4.46 |
| O (%) | 55.44 | 56.13 | 58.38 | 56.87 |
| C/N | 73.57 | 55 | 46.05 | 64.85 |
| H/C | 0.09 | 0.09 | 0.1 | 0.09 |
| O/C | 1.58 | 1.64 | 1.77 | 1.63 |

### 3.1.2 Structural analysis of FRs

The BET/BJH technique of investigation of FRs (S1-S4) as per results presents distinctly different porosity conditions, due to the previous processing type, severity undergone and fermentation-induced modifications, as shown in Table 4 and also Fig. 1(a). Results confirmed that S1 gives the highest surface area of 8.21 m²/g and a pore volume of 0.0285 cm³/g versus all FRs samples, which is due to the low severity undergone previously, in that of hydrothermal treatment (180 °C) for 30 mins, this method resulting in a strong, rigid hemicellulose surface sustaining without consistent damage being done to the lignin matrices. However, the microchemistry in the S2 showed the immense reduction in surface area (1.38 m²/g) and its corresponding pore-volume (0.00795 cm³/g) due to its conditions of high severity, in which the pore-structures underwent complete collapse owing to acid-initiated hemicellulose and repolymerized of lignin. While the S3 shows a moderate porosity (4.05 m²/g, 0.00844 cm³/g), which allows the accessible mesopores at (4.90 nm), possibly caused by the ionic ethylene-glycol solvent, which causes solvolysis of the available lignin present, operating at moderate-severity conditions achieved through higher lignin selectivity. It is noteworthy that the S4 showed extreme formation of fermentation-caused closure of the pores, and it gives a lower surface area (0.21 m²/g), a 71.4% decrease from the precursory EG2:0.735 m²/g). This was accompanied by major increases in pore width (72.06 nm) together with a decrease in pore volume (0.045 cm³/g) due to clogging, together with recondensation of lignin caused by microbial extrapolymeric substances (EPS) during photo-fermentation of the substrates [65]. The clogging of surface accessibility by biofilm formation and repolymerization of phenolics turned mesopores into macropores with transportation capability. This demonstrates that in high-severity pretreatments (EG2), although there was initially enhanced porosity, this became easily amenable to biological degradation, as opposed to those under low-severity pretreatments (HTP1), which retained structural integrity due to lower degradation [65, 66].

**Table 4:** BET and BJH Analysis of FRs

| Sample | BET (m²/g) | Pore Volume (cm³/g) | Pore Size (nm) |
|---|---|---|---|
| S1 | 8.21 | 0.0285 | 14.70 |
| S2 | 1.38 | 0.00795 | 4.37 |
| S3 | 4.05 | 0.00844 | 4.90 |
| S4 | 0.21 | 0.045 | 72.06 |

### 3.1.3 Functional groups analysis of FRs

The FTIR analysis of lignocellulosic biomass is principally of value for the qualitative analysis of the major structural components, i.e., cellulose, hemicellulose, and lignin, which highlights the structural characteristics of FRs [5]. The major absorbance peaks found are 3400-3200, 2949, 2360, 1730, 1642, 1515, 1443, 1378, 1251, 1040, 898 and 816 cm$^{-1}$ corresponding to OH stretching vibration, CH stretching vibration (alkane), O–C–O ($CO_2$), CO stretching vibration (xylan), CO stretching vibration (lignin), benzene ring stretching vibration, $CH_2$ bending vibration (cellulose), $CH_3$ bending vibration (lignin), CH bending vibration (cellulose and hemicellulose), CO stretching vibration (cellulose and hemicellulose), β-glycosidic bond vibration (cellulose), and C-N (chemisorbed N) respectively [13, 67]. There were slight differences in the absorbance peak positions and some substantial differences in the absorbance intensity of these marked peaks between samples (Fig. 1(b)). This may be due to the divergent nature of the structures and contents of cellulose, hemicellulose and lignin resulting from pretreatment and post-PFHP treatment of the substrates, which determine the thermochemical conversion of these FRs-based feedstocks [13, 28, 67].

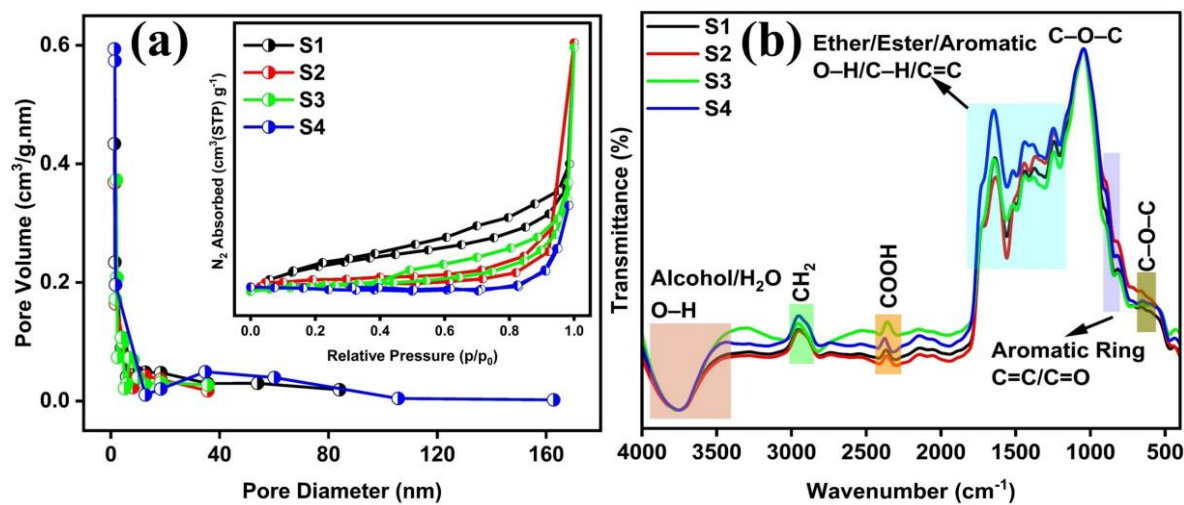

**Fig. 1:** (a) pore distribution analysis of FRs, (inset) $N_2$ adsorption/desorption isotherm, (b) FTIR spectrum of FRs

### 3.2 Pyrolysis Product Distribution of FRs

The pyrolytic products of FRs (S1–S4) followed a different pattern due to the nature of their physicochemical properties. The yields of biochar produced were fully in the range of 0.2995 to 0.374 g/g FR, with S2 (0.374 g/g FR) and S3 (0.372 g/g FR) given the higher yield of biochar (Table 5). Both samples have higher quantities of solid carbon (fixed carbon: FC: 15.82% and 15.11%) and moderate ash content (AC: 16.71%), which lends itself to carbon retention capability. The yield of S1 (0.2995 g/g FR) being the least was due to its high ash content (27.57%), resulting in a low yield of fixed carbon (6.88%). It favored the formation of more easily inorganic residues than stable carbon, while in the case of S4 (0.3223 g/g FR), a more intermediate behavior was seen due to its intermediate 22.95% ash content and VC (60.09%) [68-70]. The yield of bio-oil was maximum with S1 (42.8 ml/g FR), reinforcing that the higher total of volatile matter (61.69%) and moderate quantity of oxygen (O/C: 1.58), resulting in higher production of condensable vapors [71]. When analyzing sample S2, the bio-oil was almost as high (40.0 ml/g FR) but had its yield diminished slightly through secondary cracking at a greater temperature occupancy due to its great content of volatiles (65.85%). The least yield of bio-oil was found for S3 (36.8 ml/g FR), due in the main to its great content of oxygen (O/C: 1.77), in which the carbon was channeled mainly to gas phase products ($CO/CO_2$) via decarbonylation/decarboxylation. In contrast, the S4 showed resilience, as its lower volatility (60.09%) was compensated by efficient devolatilization and produced a total 39.6 ml/g FR bio-oil [71, 72].

The yields and compositions of the gases further highlight how the synergies were driven by the feedstock, since the total volume of gas obtained was at its maximum for the S1 (114.2 ml/g FR) and S4 (117.2 ml/g FR), where indicated that the oxygen contents were greater (O: 55.44% and 56.87%) so that these units gave rise to the production of the CO and $CO_2$ by a breakdown of the bonds. The presence of $CO_2$ (which was found to be between 35 and 37% of the total) was marked in the S3 (41.0 ml/g FR) and S4 (43.4 ml/g FR), showing their high O/C ratios (1.77 and 1.63) and carbonylation products, which the catalytic ash had formed in both these feedstocks. The greatest yield of $H_2$ and $CH_4$ was found in the S1 (23.1 ml/g $H_2$; 15.4 ml/g $CH_4$), resulting in a moderate H/C ratio (0.09) and a low nitrogen and sulphur content, which gave a series of non-hazardous products. The least gas was produced from the S2 (107.4 ml/g FR), where the high content of volatiles in the matter tended to favor the production of bio-oils, arising from its low O/C ratio (1.64), which prevents certain gasification reactions. Of particular interest is that the S3 gave upwards of 53% of its volatiles in the form of gas (47% in the form of bio-oils), and the explanation of this lies in the elevated oxygen content (58.38%) present, and the catalytic ash [73]. The syngas qualities were in favor of S1 ($H_2$/CO: 0.69), while S4 gave a steady average of all product types, which made it suitable for the purposes of its incorporation into biorefinery processes. It will be noted from these investigations that the composition of the feedstock (and especially the ash, fixed carbon and O/C ratio) is one of the factors that govern the distribution of the products of pyrolysis, for since large quantities of volatile matter favor the production of bio-oils, the oxygen contents lead to gasification, and the presence of a large fixed carbon balance, to stability of biochar [74-76].

**Table 5:** Pyrolysis Product Distribution of FRs

| Sample | Biochar (g/g FR) | Bio-oil (ml/g FR) | Gas Composition (ml/g FR) | | | | Total Gas (ml/g FR) |
|---|---|---|---|---|---|---|---|
| | | | $H_2$ | $CH_4$ | CO | $CO_2$ | |
| S1 | 0.2995 | 42.8 | 23.1 | 15.4 | 33.6 | 42.1 | 114.2 |
| S2 | 0.3740 | 40.0 | 21.7 | 14.5 | 31.6 | 39.6 | 107.4 |
| S3 | 0.3720 | 36.8 | 22.5 | 15.0 | 32.7 | 41.0 | 111.2 |
| S4 | 0.3223 | 39.6 | 23.8 | 15.9 | 34.6 | 43.4 | 117.7 |

### 3.3 Thermogravimetric Analysis

The thermochemical transformation characteristics of all of the samples were analyzed through thermogravimetric (TG) and differential thermogravimetric (DTG) methods to assess their biochar transformation characteristics, as well as to resolve the process parameters. The samples were thermally decomposed through a temperature zone of 30 °C to 700 °C with changing ramp rates of 10, 20 and 30 °C/min. The thermomechanical degradation of the FRs and their temperature association were analyzed through three stages: the loss of moisture and light volatiles (stage I), active pyrolysis (stage II), and passive pyrolysis with lignin degradation (stage III). Stage I analyzed the losses to an average of 4.38%, 5.15% and 6.10% mass loss with ramp rates of 10, 20 and 30 °C/min, respectively, from 30 to 150 °C, resulting primarily from moisture and light volatiles [77]. The mass loss for stage II was 53.29%, 51.67% and 50.64% respectively, with average decomposition rates of -0.89%/°C, -0.72%/°C, and -0.63%/°C, respectively, for the ramp rates of 10, 20 and 30 °C/min. The higher decomposition rates in stage II can be related to the thermal degradation of hemicellulose and cellulose. The decomposition of the high-molecular-weight hemicellulose and cellulose in this stage would give off the most volatile matter, and thus allow for the collection of the pyrolysis by-products (bio-oil and pyrolysis gas) [78]. The average mass loss for stage III was 8.84%, 9.44% and 10.71% for ramp rates of 10, 20 and 30 °C/min, respectively, with average decomposition rates of -0.10%/°C, -0.09%/°C and -0.12%/°C, respectively, for the different ramp rates. The lower average decomposition rates in stage III are due to the high thermal stability of the lignin, which will only degrade at the higher temperatures [77, 78]. As seen in Fig. 2(a, b, c) and Fig. 2(d, e, f), there is a right-hand shift of the DTG and the TG curves with increasing ramp rates from 10 °C/min to 30 °C/min in stage II. This would appear to be from the thermal lag associated with the heat transfer limitations at the higher heating rates, resulting in the higher pyrolysis temperatures for thermal degradation of the FRs [79]. The increasing heating rate had no correlation with stage I, indicating that the thermal evaporation for the water volatiles does not appear to be affected by ramp rate [80]. Beyond temperatures of 650 °C, minimal weight loss is seen, indicating that the thermochemical conversion of the FRs is complete. The residual weights (biochar) at the end of the pyrolysis of the FRs at 700 °C for the two FRs is (S1: 34.45%, S2: 32.53%, S3: 31.82%, S4: 35.15%), (S1: 36.27%, S2: 33.49%, S3: 31.95%, S4: 33.23%) and (S1: 34.65%, 33.83%S2: 28.42%, S4: 33.29%) for the ramp rates of 10, 20 and 30 °C/min. Through the TG analysis, it can be seen that for the S3 sample, there is the least amount of residual biochar across all ramp rates, seemingly implying a higher degree of volatile compounds (VC). At the same time, the S1 sample gave the lowest biochar yield, indicating that the volatile compounds were absent in lower quantities. Through a DTG, the highest decomposition rate at the elevated temperatures was again said to be for the S3 sample, followed by S4, with S2 giving intermediate results for both the TG and DTG, remaining closely associated with the averaged data of all the samples [13, 81]. The complexity of the FRs has shown subsequently results that vary in contrast to other studies [82], which relate different results in the measuring of the different TG and DTG results for the different biomass types.

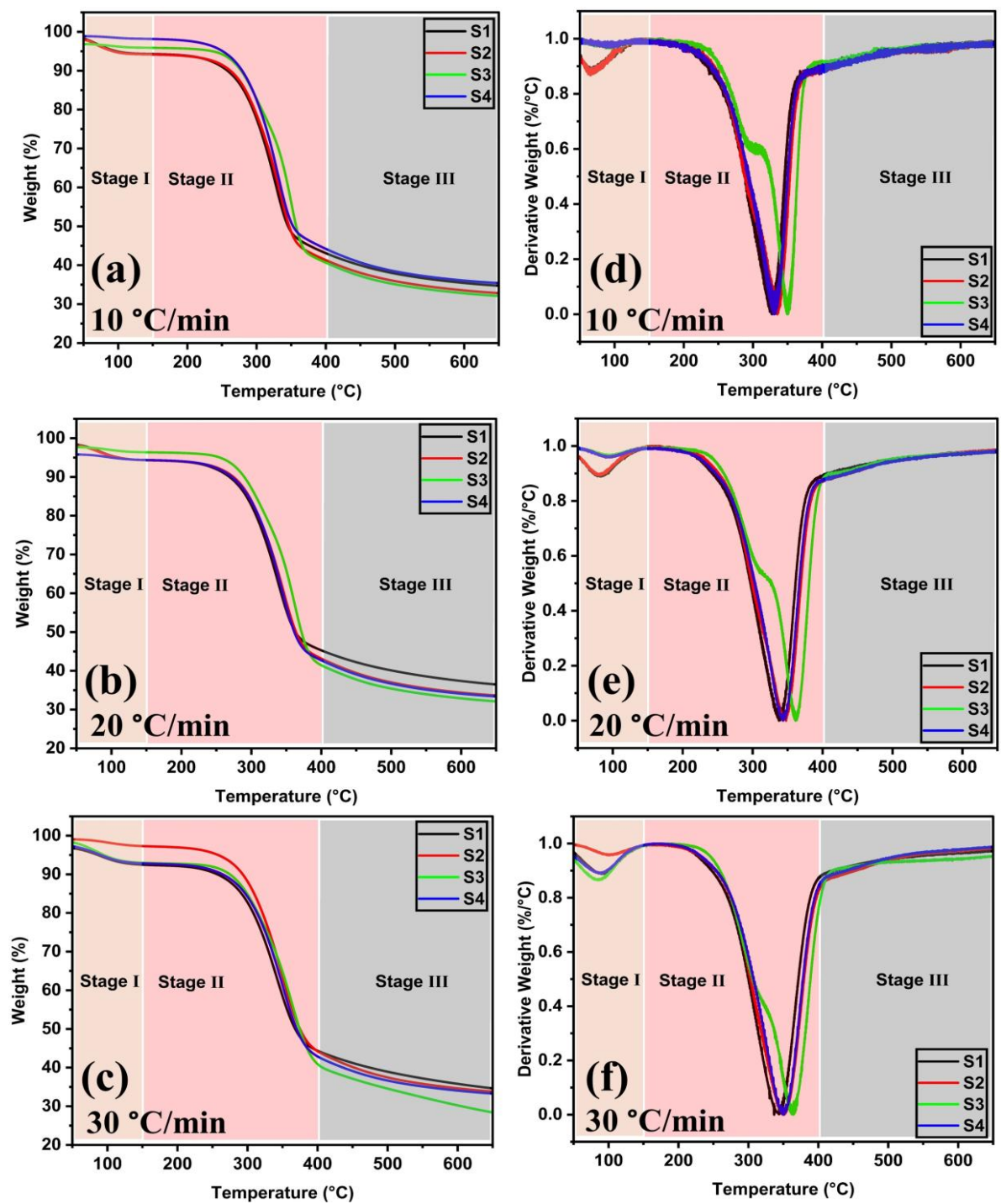

**Fig. 2:** TG curve at (a) 10 °C/min, (b) 20 °C/min and (c) 30 °C/min ramp rate, and DTG curve at (d) 10 °C/min, (e) 20 °C/min and (f) 30 °C/min ramp rate.

### 3.4 Iso-conversional methods for kinetic analysis

The thermo-kinetic parameters of biomass are crucial for the efficient design of thermochemical processes, as pyrolysis involves a multistep reaction mechanism with complex reactions. In this study, the pyrolysis kinetics of S1, S2, S3, and S4 are used to understand pyrolysis behaviors and serve as a reference for the process. Least squares fit plots were generated to assess the relationship between 1/T vs ln($\beta$), 1/T vs ln($\beta/T^2$), and 1/T vs ln($\beta/T^{1.92}$) at conversion rates from 0.1 to 0.9, increasing in 10% conversion for each step, using the FWO, KAS, and Starink models. As shown in Fig. 3(a-l), the lines are nearly parallel across all samples S1-S4, indicating that all three Ea calculation methods are reliable [13, 83].

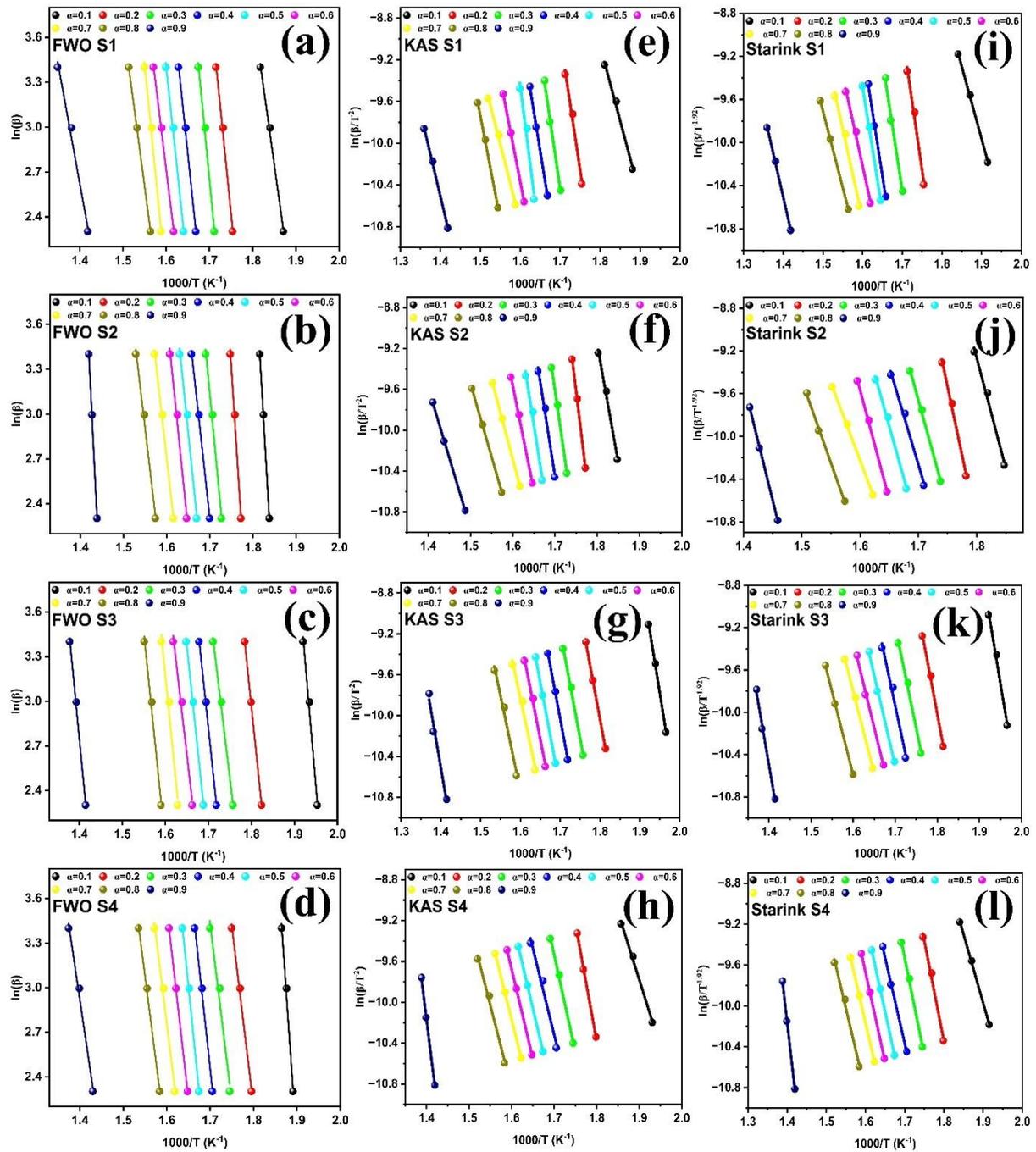

**Fig. 3:** (a, b, c, d) ln$\beta$ vs 1/T plot at different $\alpha$ for Ea calculation using FWO model, (e, f, g, h) ln($\beta/T^2$) vs 1/T plot at different $\alpha$ for Ea calculation using KAS model, (i, j, k, l) ln($\beta/T^{1.92}$) vs 1/T plot at different $\alpha$ for Ea calculation using Starink model.

The three methods produce similar Ea trends between 0.1 and 0.9 conversions with a step size of 0.1. The Ea values of S1 were estimated to range from 125.56 to 237.54 kJ/mol, 119.51 to 250.24 kJ/mol, and 111.92 to 211.33 kJ/mol with an average of 196.80 kJ/mol, 175.85 kJ/mol, and 160.06 kJ/mol by FWO, KAS, and Starink models, respectively (Fig. 4(a). As shown in Fig. 4(c, d, e), the Ea values changed significantly with the increase of α, as from α = 0.1–0.3, the activation energies were found to increase significantly. High Ea indicates that some highly endothermic reaction occurred at this conversion point, while a decline of Ea shows the gradual transition of reaction mechanism during the pyrolysis process of fatty acids and proteins [13]. The Ea values of S2 were estimated to range from 193.47 to 460.14 kJ/mol, 114.49 to 282.64 kJ/mol, and 120.61 to 214.45 kJ/mol with an average of 277.80 kJ/mol, 187.23 kJ/mol, and 162.97 kJ/mol by FWO, KAS, and Starink models, respectively. The Ea values of S3 were estimated to range from 189.02 to 266.60 kJ/mol, 151.45 to 201.30 kJ/mol, and 130.08 to 199.45 kJ/mol with an average of 221.95 kJ/mol, 172.56 kJ/mol, and 157.68 kJ/mol by FWO, KAS, and Starink models, respectively. The Ea values of S4 were estimated to range from 156.38 to 333.39 kJ/mol, 110.2 to 278.09 kJ/mol, and 111.91 to 277.87 kJ/mol with an average of 209.10 kJ/mol, 160.66 kJ/mol, and 157.43 kJ/mol by FWO, KAS, and Starink models, respectively. All samples with all three methods showed the linear correlation coefficients for determining the Ea ranged from 0.96 to 0.99, expressing the reliability of the calculation methods and the given data [13, 83].

Furthermore, all samples highest Ea (S1: 237.54 kJ/mol, S2: 460.14 kJ/mol, S3: 266.60 kJ/mol and S4: 333.39 kJ/mol) were much lower than the Ea of previous study on fermentation residues of pine sawdust which shows highest Ea of 853.13 kJ/mol, 789.19 kJ/mol, and 862.87 kJ/mol attained by KAS, FWO, and Starink method, respectively [13]. This drastically lowered Ea of S1-S4 suggests its better pyrolytic conversion with lower Ea requirement, a suitable candidate for further utilization in multiple applications. Ea values at the beginning are considered low due to the cleavage of containing weak bonds (e.g., O-H) and the elimination of volatile components from the FR's surface. After α = 0.1, the second pyrolysis stage occurs, accompanied by the generation of various volatiles, resulting in different rates of volatile escape. The volatiles evolved to cover the surface of FRs and form carbonaceous char, which acts as a barrier insulating the heat from reaching the surface below, resulting in a decrease in reaction activity. Ultimately, more energy is required to activate the reaction, which is evident by the increased Ea with incremental conversion. Furthermore, a sudden rise in Ea at α = 0.9 was observed, which was due to deep degradation of lignin and the formation of carbon in a multi-aromatic ring system [13, 83]. Overall, the average Ea of S1 was the lowest among all, which was due to S1's higher cellulose content compared to lignin; however, in S2-S4, a higher lignin content remained, as they produced more $H_2$ during fermentation, utilizing exposed cellulose, and the remaining FRs contained a higher lignin content. In future research, focusing on catalytic or co-pyrolysis may be a suitable way to reduce the pyrolysis activation energy.

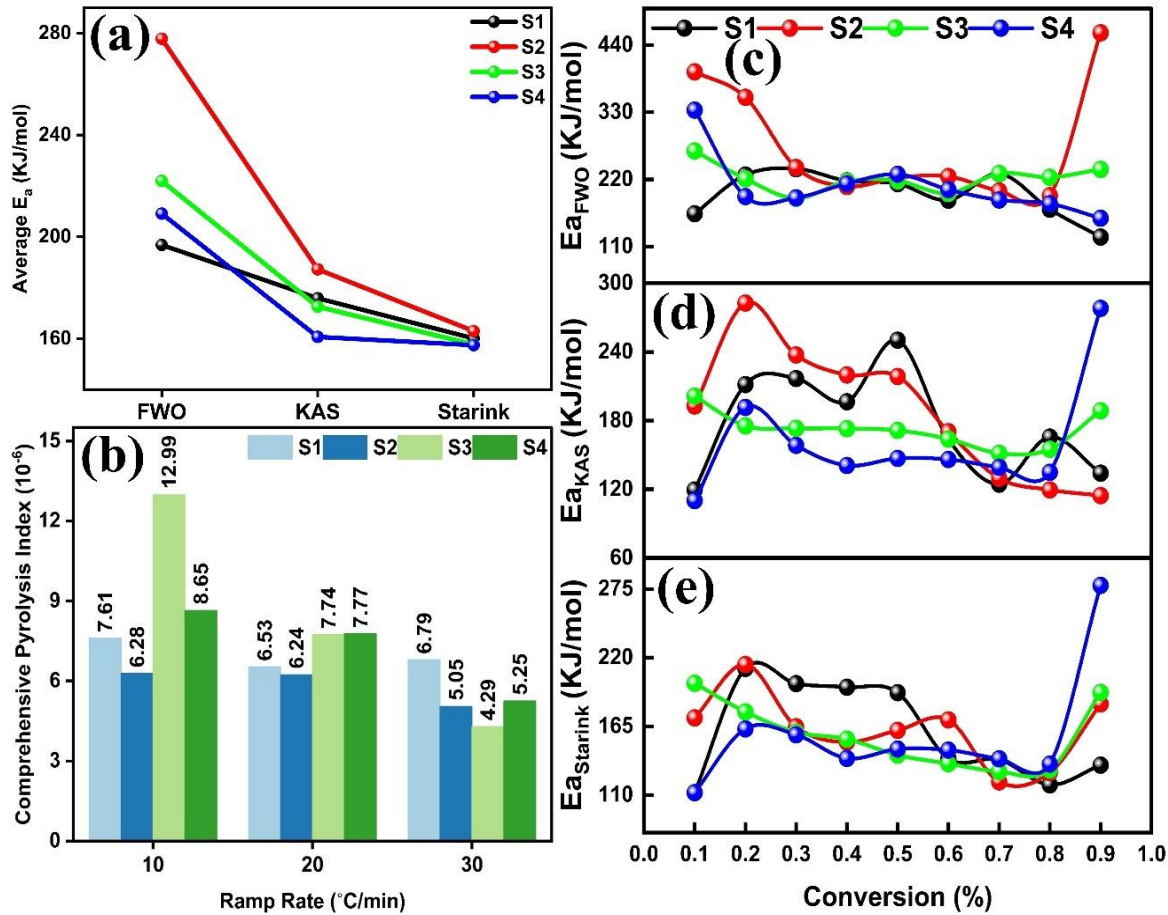

**Fig. 4:** (a) Avg Ea of all samples calculated using FWO, KAS, and Starink model, (b) Comprehensive pyrolysis index of all samples, (c) Ea of all samples calculated at different α using FWO model, (d) Ea of all samples calculated at different α using KAS model, (e) Ea of all samples calculated at different α using Starink model.

### 3.5 Correlation between FRs characteristics and comprehensive pyrolysis index

Based on the Comprehensive Pyrolysis Index (CPI) data for samples S1-S4 across heating rates of 10, 20, and 30°C/min, distinct reactivity patterns emerge that align with compositional and kinetic properties detailed in the manuscript Fig. 4(b). The S3 consistently demonstrates superior pyrolysis efficiency at 10°C/min with the highest CPI = 1.30x10-5), due to its high oxygen-content matrix (O/C=1.77), which encourages aggressive decarbonylation processes and devolatilization processes driven by catalytic ash, thus optimizing the occurrence of volatiles, despite its moderate volatile yield (62.94%). This level of initial reactive ability shows quickly with increases at heat-up rate (CPI = 7.74x10-6 @20 °C/min; 4.29x10-6 @30

°C/min), where thermal lag dominates reaction reactivity, in conjunction with the gas-shift production achieved, where 53% of volatiles were converted into gases. While the sample S4 shows a moderate yet stable level of performance at all heat-up rate regimes (8.65x10-6 @10 °C/min; 7.77x10-6@20 °C/min; 5.25x10-6@30 °C/min), which corresponds with its balanced carbon-oxygen composition (C:34.96%, O:56.87%) together with its graphitic domains resulting from ethylene glycol pretreatment, which add resistance to decomposition from thermal treatment, and enable progressive biochar-energy co-production [84, 85]. Naturally, S1 shows a high level of certainty of reactivity (7.61x10-6@10 °C/min; 6.53x10-6@20 °C/min; 6.79x10-6@30°C/min), which down to its high ash content (27.57%), which mitigates the effect of heating and mobility of volatiles, thus resulting in later peaks of temperature of decomposition, and intensity of DTG outputs.

Meanwhile the S2 gives a promising performance on the first instance at the 10 °C/min cycle (6.28×10-6) due to its high volatile matter content (65.85%), however it also suffers the fastest fall in CPI (5.05 × 10-6) @30 °C/min, due to the induced cracking products as a result of the accelerated heating, which diminishes the bio-oil produced in the early stages and gives rises to the quantity of non-condensable gases [84]. The reduction of all samples with this reactivity is tangible through the decrease in CPI values together with an increase in heat-up rates, demonstrating a consistent trend through its thermal lag effect. The steepest reactivity, however, is in S3, which shows a heightened oxygen concentration to vapor bonding, while S4 demonstrates good resilience to show stability through structure. These dynamics reinforce the feedstock hierarchy established in kinetic analyses: oxygen content governs peak reactivity, ash content imposes diffusion limitations, and volatile-ash balance determines process scalability, positioning S3 for maximal bioenergy recovery in moderate-temperature reactors and S4 for robust industrial applications requiring heating-rate flexibility [84, 86].

### 3.6 Thermodynamic parameter analysis

The comprehensive thermodynamic analysis of fermentation residues (S1–S4) paints a clear picture of how conversion-dependent variations across the three model-free methods (FWO, KAS, Starink) reveal the underlying complexity of biomass pyrolysis. The average values of pre-exponential factor (A), enthalpy changes (ΔH), Gibbs free energy (ΔG) and entropy change ΔS are given in Table 6, while the detailed discussion of these thermodynamic parameters including all conversion points from 0.1- 0.9 and ramp-rate-specific averages, provided in the supplementary file. Starting with the A, which we calculated using the Coats-Redfern method based on activation energies from each model, the values spanned extraordinary ranges: $1.02 \times 10^7$–$2.43 \times 10^{19}$ s$^{-1}$ for S1, $7.87 \times 10^{13}$–$2.77 \times 10^{36}$ s$^{-1}$ for S2, $1.48 \times 10^{15}$–$3.32 \times 10^{25}$ s$^{-1}$ for S3, and $3.10 \times 10^9$–$2.08 \times 10^{31}$ s$^{-1}$ for S4, with method averages of $3.73 \times 10^{18}$, $3.08 \times 10^{35}$, $3.69 \times 10^{24}$, and $2.31 \times 10^{30}$ s$^{-1}$ via FWO, respectively. These magnitudes, which far exceed the $10^9$ s$^{-1}$ threshold, signal complex multi-step reaction networks rather than simple surface processes [87], directly reflecting each feedstock's unique structural and compositional fingerprint [13].

Moving to the ΔH, S2 consistently demanded the highest energy input averaging 272.7 kJ/mol (FWO), 182.2 kJ/mol (KAS), and 157.9 kJ/mol (Starink) peaking at 391.5 kJ/mol at α=0.1 before plunging to 108.8 kJ/mol by α=0.9, a pattern that mirrors pine sawdust pyrolysis where high volatile matter (65.85%) drives substantial glycosidic bond scission [13]. Conversely, S4 required the least enthalpy (204.1-152.4 kJ/mol) thanks to ethylene glycol pretreatment that fragmented its structure, while S1 and S3 fell in the moderate range (191.7-216.9 kJ/mol). Every sample showed a universal ΔH decline beyond α=0.5, marking the shift from endothermic volatilization to exothermic carbonization [13, 83, 87, 88]. Similarly, the ΔG revealed the thermodynamic reactivity: S4 faced the steepest barrier (269.9-271.3 kJ/mol) due to graphitic domains formed during harsh pretreatment, while S1 show the lowest resistance (176.2-177.6 kJ/mol) from its cellulose-rich composition. Interestingly, ΔG increased steadily with conversion for all samples, with S4 climbing most dramatically from 230.5 to 323.7 kJ/mol as lignin decomposition dominated late-stage pyrolysis [13], reflecting the growing molecular complexity of aromatic condensation. Entropy (ΔS) perhaps told the most vivid story: S2 swung wildly from +387.0 kJ/mol·K at α=0.1 to -47.5 kJ/mol·K at α=0.8, capturing the chaotic burst of devolatilization followed by rapid biochar ordering, while S4 remained stubbornly negative throughout (-93.7 to -198.4 kJ/mol·K), confirming stable graphitization [13, 87]. The oxygen-rich matrix of S3 (O/C=1.77) drove consistently negative ΔS (-18.3 to -136.0 kJ/mol·K) via decarbonylation, whereas S1 showed transitional behavior (36.3 to -60.6 kJ/mol·K).

Critically, every sample crossed from positive to negative ΔS beyond α=0.5, a universal signature of the shift from disorder-driven volatile release to ordered carbon matrix stabilization governed by lignin's thermal resilience (180-900°C) [13, 83, 87, 88], consistent with bamboo pyrolysis where declining ΔS indicated enhanced structural order [83]. Methodologically, FWO consistently overestimated ΔH and ΔS relative to KAS and Starink a known bias in integral approximations, yet all three models agreed on the fundamental hierarchies: S2 > S3 > S1 > S4 for ΔH, and S4 > S3 > S2 > S1 for ΔG. These variations underscore the precision limitations of individual model-free approaches and reinforce the need for multi-model validation in complex biomass systems [42]. Ultimately, these thermodynamic signatures directly correlate with composition: S2's volatility maximized enthalpy and entropy fluctuations, S4 pretreatment minimized ΔH while maximizing ΔG, S3 O/C ratio drove negative entropy, and S1 high ash content (27.57%) suppressed enthalpy at high conversion. The complete conversion-point data, method-specific averages, and ramp-rate variations are fully documented in the supplementary file, providing the granular foundation for these interpretations and highlighting why robust biomass kinetic characterization demands multiple analytical perspectives, much like the inverse ΔG-spontaneity relationships observed in Prosopis Juliflora biochar studies [89].

**Table 6:** Thermodynamic parameters of all FRs during pyrolysis

| Parameter | Model | S1 | S2 | S3 | S4 |
|---|---|---|---|---|---|
| Pre-exponential factor (A) sec$^{-1}$ | FWO | 3.73E+18 | 3.08E+35 | 3.69E+24 | 2.31E+30 |
| | KAS | 3.65E+18 | 2.86E+23 | 8.08E+17 | 5.20E+17 |
| | Starink | 3.51E+16 | 1.39E+17 | 5.27E+17 | 4.99E+17 |
| Enthalpy (ΔH) kJ/mol | FWO | 191.66 | 272.72 | 216.94 | 204.04 |
| | KAS | 170.71 | 182.18 | 167.57 | 155.62 |
| | Starink | 154.94 | 157.92 | 152.70 | 152.39 |
| Gibbs Free Energy (ΔG) kJ/mol | FWO | 176.18 | 204.32 | 234.62 | 269.88 |
| | KAS | 177.60 | 206.38 | 236.28 | 270.96 |
| | Starink | 177.43 | 206.86 | 236.20 | 271.29 |
| Entropy (ΔS) kJ/mol·K | FWO | 36.29 | 121.54 | -18.32 | -93.66 |
| | KAS | -8.41 | -33.81 | -111.07 | -192.30 |
| | Starink | -33.84 | -78.21 | -135.97 | -198.36 |

### 3.7 Coats Redfern Method for internal pyrolysis mechanism investigation

The results collected through Coats Redfern's method expand on the pyrolysis mechanisms present within FRs (S1 to S4). There is certainly no doubt that the mechanisms are governed by controlled diffusion processes, thus proved by a multi-parameter study [44]. For sample S1, the diffusion model exhibiting cylindrical symmetry and thus designated as D4 has evolved as the most likely, with an activation energy of 159.62 kJ/mol (Table 7). This model is consistent with the high ash content present within the sample S1 (27.57%), where the minerals create tortuous means of diffusion pathways [44, 90], exacerbated by the low quantity of volatile matter (61.69%) and fixed carbon (6.88%), thus limiting the mobility of the volatiles. The fact that the model is consistent can be seen when looked at with the KAS-derived activation energy present (175.85 kJ/mol) and Gibbs free energy thermodynamic (ΔG = 177.4 kJ/mol). This suggests a measure of energy performance on the energetics present, which accumulate once the volatiles are released. For sample S2, this conforms to spherical diffusion (D3), with the evidence showing that this sample has the highest quantity of volatiles (65.85%). The sample will thus create internal gradients of pressure during the decomposition steps; thus, it needs radial diffusion for the volatiles themselves to spontaneously diffuse through the developing char [13, 44, 90]. The connection of the spherical symmetry can also be viewed in the concentric layers of structural devolatilization that are observed, while the fact that there are negative entropic transitions (-47.5 kJ/mol·K at α=0.8) is indicative of the transformation from chaos to order of the process which this model describes. The values of activation energy derived from CRM are seen to be bracketed by those derived from FWO (277.80 kJ/mol) and Starink (162.97 kJ/mol), thus further proving that similar mechanisms are present.

The sample S3 shows this common model (D3, Ea = 192.17 kJ/mol) because oxygen causes cross-linking, leading to an O/C ratio of 1.77. The significant amount of oxygen leads to the production of stable ether or carbonyl links that restrict isotropic diffusion capabilities that are dictated by changes in pore size (4.90 nm S3 vs 14.70 nm S2) [13, 44, 90]. The decline in the enthalpy (ΔH: 262.3→182.6 kJ/mol via FWO) is a reflection of the decarbonylation pathways exerted on the volatiles that again confirm oxygen is sought as advantageous in many products, while the value of A ($10^{11}$ $s^{-1}$) implies solid-state diffusion in excess throughout the reaction pathway. Validation is sought through correlation of KAS Ea (172.56 kJ/mol) and Starink (157.68 kJ/mol) Ea values, which confirm similar energy barriers of activation that are geometrically advantageous through the toughened matrix of oxygen. Finally, for sample S4, this follows the D3 method (Ea = 218.05 kJ/mol) once again. Enabled by the development of graphite phases resulting from treatment with the substance ethylene glycol prior to activation. Thus, with the lowest surface area (0.386 $m^2$/g) and enlarged pore radius (22.15 nm), this allows for thick layers to be produced, which illustrates that the resultant char developed retains a layered structure, causing extensive radial diffusion effects before the movement of the volatiles [13, 44, 90]. The highest value of the CRM finding of Ea corresponds with the trends of FWO (209.10 kJ/mol) as well as with Starink (157.43 kJ/mol), while ΔG value is significantly elevated (271.3 kJ/mol) thus dictating to some extent barriers predominate energy circulation to disrupt the π stacked geometry of the graphitic layers present in the bioactive char.

**Table 7:** Calculated Ea of FRs samples using Coats-Redfern Method at different heating rates

| Heating Rate (°C/min) | Model | S1 | | S2 | | S3 | | S4 | |
|---|---|---|---|---|---|---|---|---|---|
| | | $R^2$ | Ea (KJ/mol) | $R^2$ | Ea (KJ/mol) | $R^2$ | Ea (KJ/mol) | $R^2$ | Ea (KJ/mol) |
| 10 | O1 | 0.77 | -117.28 | 0.76 | -125.34 | 0.67 | -81.80 | 0.70 | -88.86 |
| | R3 | 0.47 | -2.71 | -0.12 | -0.28 | 0.62 | -2.63 | 0.56 | -2.24 |
| | D1 | 0.81 | 85.32 | 0.97 | 122.85 | 0.98 | 88.84 | 0.99 | 95.38 |
| | D2 | 0.84 | 96.16 | 0.98 | 134.48 | 0.98 | 96.96 | 0.99 | 104.12 |
| | D3 | 0.45 | 4.87 | 0.89 | 9.17 | 0.44 | 4.11 | 0.57 | 5.00 |
| | D4 | 0.85 | 100.79 | 0.98 | 139.10 | 0.98 | 100.16 | 0.99 | 107.56 |
| | A2 | 0.89 | 57.69 | 0.99 | 76.72 | 0.96 | 53.73 | 0.98 | 58.00 |
| | A3 | 0.89 | 57.69 | 0.99 | 76.72 | 0.96 | 53.73 | 0.98 | 58.00 |
| 20 | O1 | 0.77 | -117.23 | 0.75 | -119.05 | 0.67 | -82.81 | 0.75 | -92.89 |
| | R3 | 0.09 | -0.79 | 0.07 | -0.74 | 0.62 | -2.69 | 0.70 | -2.51 |
| | D1 | 0.99 | 121.04 | 0.98 | 119.30 | 0.99 | 90.46 | 0.96 | 92.47 |
| | D2 | 1.00 | 132.27 | 0.99 | 130.50 | 0.98 | 98.70 | 0.96 | 101.26 |
| | D3 | 0.86 | 8.46 | 0.86 | 8.41 | 0.43 | 4.16 | 0.63 | 4.70 |
| | D4 | 1.00 | 136.70 | 0.99 | 134.94 | 0.98 | 101.94 | 0.97 | 104.74 |
| | A2 | 1.00 | 74.87 | 1.00 | 74.09 | 0.96 | 54.63 | 0.96 | 56.55 |
| | A3 | 1.00 | 74.87 | 1.00 | 74.09 | 0.96 | 54.63 | 0.96 | 56.55 |
| 30 | O1 | 0.72 | -82.40 | 0.66 | -78.48 | 0.67 | -80.20 | 0.77 | -117.90 |
| | R3 | 0.81 | -3.28 | 0.71 | -3.18 | 0.73 | -3.10 | 0.35 | -1.12 |
| | D1 | 0.99 | 85.33 | 0.98 | 85.61 | 0.98 | 85.53 | 0.98 | 115.06 |
| | D2 | 0.99 | 93.28 | 0.98 | 93.41 | 0.98 | 93.42 | 0.99 | 125.98 |
| | D3 | 0.43 | 3.31 | 0.31 | 3.32 | 0.37 | 3.44 | 0.88 | 7.78 |
| | D4 | 0.99 | 96.41 | 0.98 | 96.48 | 0.98 | 96.53 | 0.99 | 130.30 |
| | A2 | 0.98 | 51.42 | 0.95 | 51.38 | 0.96 | 51.53 | 1.00 | 71.42 |
| | A3 | 0.98 | 51.42 | 0.95 | 51.38 | 0.96 | 51.53 | 1.00 | 71.42 |

Mechanistic validation transcends mathematical fitting through four interlocked phenomena: First, pre-exponential factors ($10^9$–$10^{14}$ s$^{-1}$) exceed the chemical reaction thresholds (<$10^9$ s$^{-1}$), confirming the complexity of diffusion. Second, DTG peak shifts illustrate delayed decomposition kinetics inherent to diffusion-limited systems. Third, BET surface area inversely correlates with CRM Ea (S1: 8.21 m$^2$/g vs. 159.62 kJ/mol; S4: 0.386 m$^2$/g vs. 218.05 kJ/mol), highlighting structural control. Fourth, thermodynamic synergy is evident: early positive ΔS (e.g., S2: +387 kJ/mol·K at α=0.1) confirms volatile entrapment, while late negative ΔS (S4: −198 kJ/mol·K) denotes graphitic ordering that impedes diffusion. The rejection of non-diffusion models (e.g., first-order O1, phase-boundary R3, nucleation A2/A3) is grounded in physicochemical incompatibility.

Unreasonably low values of Ea calculated for O1 would contradict the energy required for bond-cleavage in any cross-linked residues [13]. While cross-linking R3 models do not account for any internal diffusion gradients, relative to the broadening of the DTG peak. In contrast, the Nucleation models cannot account for the rates of Ea, which are dependent on heating rates, since steps that are rate-limiting in the various nucleation mechanisms are thermally insensitive [13, 90]. Fermentation caused structuring due to densification, occlusion of the pores, and a loss of macro-porosity, which yields barriers to transport, which would always correlate with the diffusion theory as being the mechanism of control for diffusion processes; therefore, diffusion processes being three-dimensional cumulatively account for pyrolysis processes in these systems universally [13, 65].

### 3.8 Properties of Biochar Produced from Fermentation Residues

The fermentation residues were converted into four unique biochars through pyrolysis at 700 °C. The biochar morphologies and architectures were directly related to the composition of the respective precursors, establishing beneficial catalysts for the closed-loop system.

### 3.8.1 Microstructural Transformation

Scanning electron microscopy images demonstrated how the fingerprint from each feedstock dictated the morphology, as the sample BC1 has an irregular surface with fragmented pieces and macropores larger than 36 nm (Fig. 5(a)). The composition of BC1 had a high ash percentage (27.57%), which disrupted the biochar structure and forced the release of volatiles through relatively tortuous routes associated with the minerals added to the formulation, confirming the cylindrical diffusion mechanism (D4) established in the kinetic study [91]. The sample BC2 exhibited a honeycomb structure of vesicular spherical pores approximately 20 nm in size (Fig. 5(b)). The structure of BC2 was formed as it underwent rapid devolatilization and, as volatile organics were released from the precursor (VC: 65.85%), pressure internally increased and moisture was redistributed. The previously mentioned mechanisms of internal pressure gradients and pore arrangement prevented an efficient orientation of pore structure, lowering opportunities for surface area [92, 93]. Interestingly, structure BC3 developed a relatively uniform ultra-microporous structure with an abundance of densely packed pores of less than 4 nm and smooth edges (Fig. 5(c)).

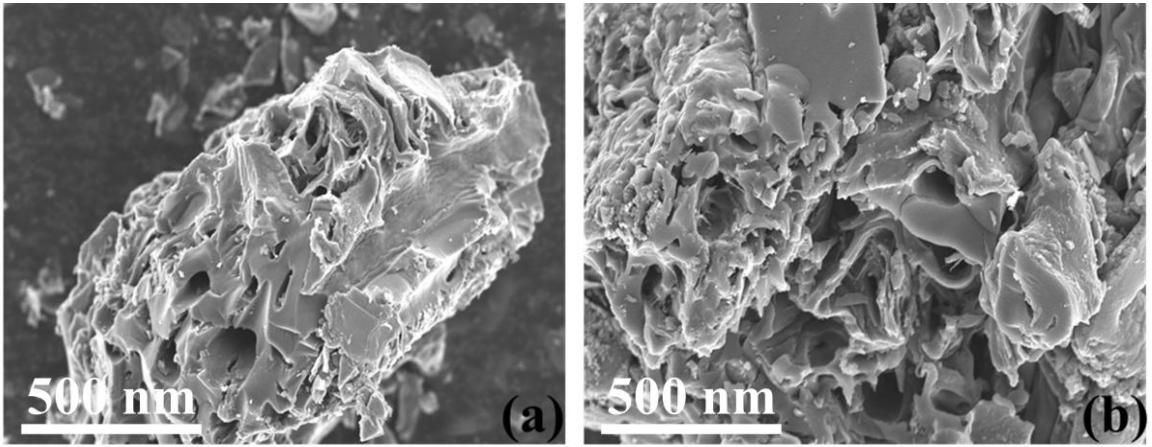
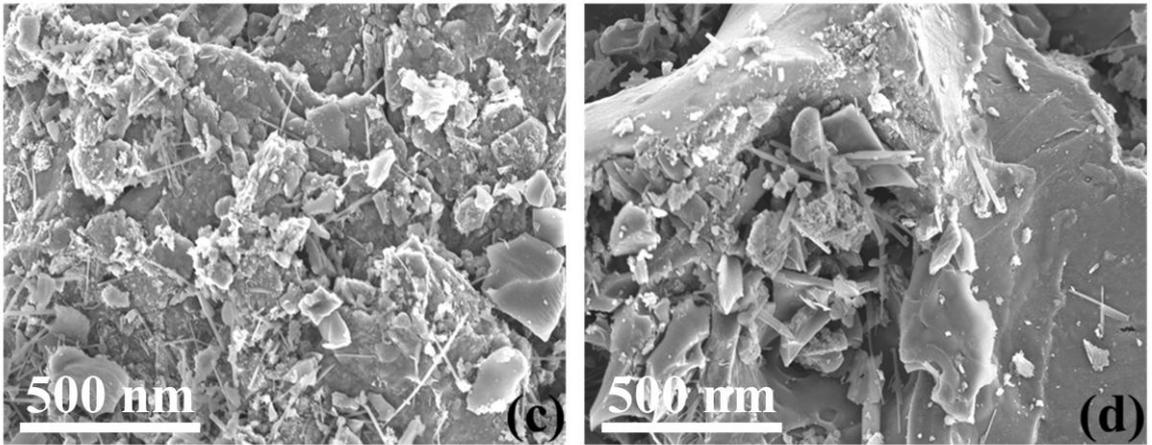
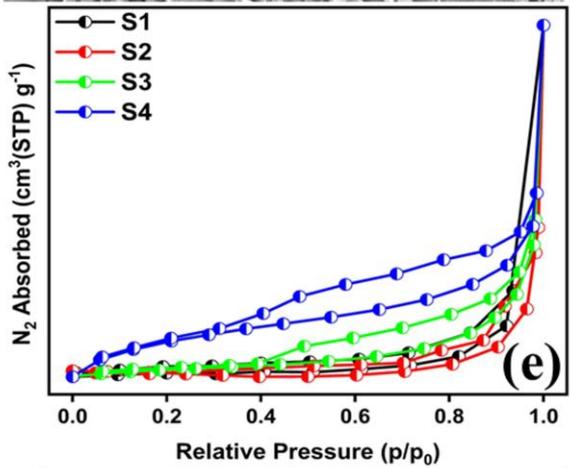
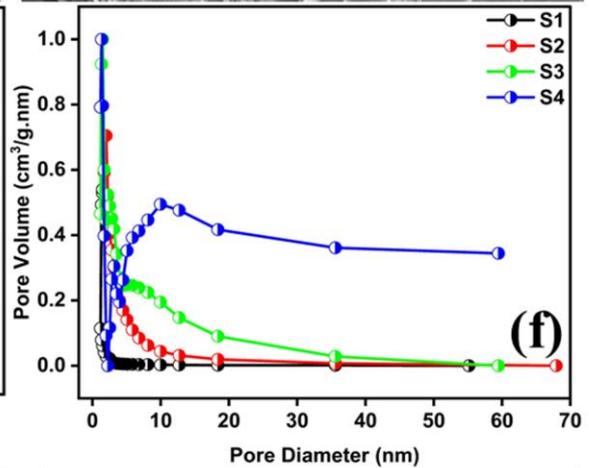
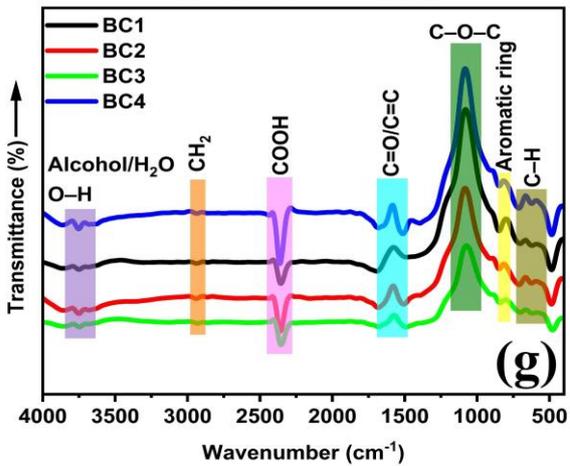
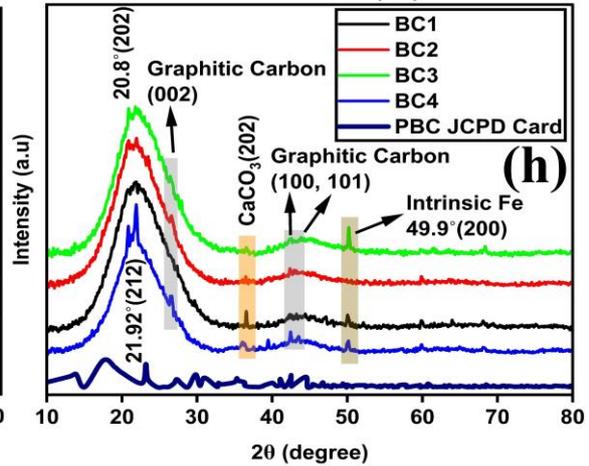

**Fig. 5:** Characteristics of biochar samples (a) SEM of BC1, (b) SEM of BC2, (c) SEM of BC3, (d) SEM of BC4, (e) $N_2$ sorption/desorption isotherm of all BC, (f) pore size distribution of all BC, (g) FTIR of all BC, (h) XRD of all BC.

The rigidity of the porous structure was caused by a cross-linking reaction mediated by $O_2$ during the decarbonylation of the precursor, preserving the internal structure, and enabling properties to sustain large BET values [94, 95]. The structure of BC4 had the most ordered architecture, demonstrating stacked graphitic domains approximately around 4 nm slit shaped mesopores that formed aligned channels in between carbon platelets (Fig. 5(d)), found because of the effect of ethylene glycol pretreatment that permitted and promoted radial diffusion, and electron conduction associated with direct interspecies electron transfer [96]. The differences in morphology were also portrayed in the quantifiable surface area and pore size distribution, as by examining the BET and BJH results respectively (Fig. 5(e, f)). The BC4 had the 2$^{nd}$ highest area (76.58 m$^2$/g) associated with the order of structure due to the alignment of the carbon domains; whereas, BC3 had a high level of oxygen associated with the matrix, and exhibited mesoporosity based on the BET result (185.14 m$^2$/g, 4.67 nm pores), a dramatic increase of 14.6 times from the damaged fermentation precursor (Table 8). BC1 and BC2 remained relatively low (7.20 and 5.98 m$^2$/g, respectively) with regard to BET area due to effects of ash and volatile-driven collapse; although, the vesicular properties of BC2 do present open access pathways.

**Table 8:** BET/BJH analysis of all FRs-based biochar

| Sample | BET (m$^2$/g) | Pore Volume (cm$^3$/g) | Pore Size (nm) |
|---|---|---|---|
| BC1 | 7.20 | 0.0129 | 6.74 |
| BC2 | 5.98 | 0.031 | 20.15 |
| BC3 | 185.14 | 0.049 | 4.67 |
| BC4 | 76.58 | 0.054 | 4.31 |

### 3.8.2 Surface Chemistry and Crystallography

Fourier-transform infrared (FTIR) spectroscopy demonstrated that oxygen-functionalization was evident for all biochar samples. However, the signatures of functionalization were more distinctive among all FRs-based BCs (Fig. 5(g)). The BC3 exhibited the greatest intensity of FTIR spectrum, distinguishing functional groups of carboxylic acid (–COOH, 2359 cm$^{-1}$) and a peak signal of carbonyl (C=O, 1575 cm$^{-1}$), resulting in acidic groups that would buffer pH and chelate cations. This peak in acidic quality fractional (574 mL) relates directly to its peak cumulative yield of hydrogen [97]. The ether stretching bond (C–O–C, 1080 cm$^{-1}$) was greatest in intensity for BC4, and since the ether bond relates to the graphitic domains, did demonstrated the capacity to shuttle electrons rapidly through biochar. BC4 also had the greatest hydrogen production rate (14.91 mL/h) [96]. Aromatic carbon (C)-hydrogen (H) bands for BC2 demonstrated that functional groups (lignin-derived redox) scavenged dissolved oxygen (even though their value reached low oxidation-reduction potential, –521 mV). The peak at aliphatic C–H signals for BC1 was detected with oxygen functionality peaks diminished, correlating to the high ash occluded instance (27.57%) structure yielding lowest performance (380mL $H_2$) [98-100].

X-ray diffraction gave additional support to crystallographic signatures and assuring important trends, as a clear dominance of minerals was displayed for BC1 which can be observed from Fig. 5(h). The three minerals of quartz peaks (20.8°, 26.64°), calcite (36.49°), dolomite (39.46°) covered the amorphous carbon signal and affirmed suppressed ordering mediated by ash [91]. A broad turbostratic carbon peak correlated to kaolinite, consistent with maximum collapse of structure induced by volatiles for BC2. BC3 evidenced intense (002) reflection and reflections at (100) signals, suggesting that oxygen-mediated cross-linking enhanced turbostratic ordering [101]. For the highest crystallinity, BC4 had sharp graphitic peaks and higher-order reflections indicative of the alignment of domains caused by ethylene glycol that enabled efficient electron transfer [96, 102].

The derived hierarchy (BC4 > BC3 >> BC1 ≈ BC2) emphasizes that precursor composition is the primary determinant of function over initial porosity. BC3's microporous and oxygen-rich framework sustained the highest total hydrogen yield because of its extended buffering capacity of pH and improved microbial habitat; BC4 had graphitic mesopores that facilitated electron transfer at the highest rates of production [103]. BC1 is macro-porous with a limiting amount of ash, and BC2 is structurally degraded (collapsed) with less defined porosity, leading to reduced efficiency. This is an example of utilizing the cost-effective precursor chemistry, crystallographic evolution, engineered porosity, and the "three-birds-with-one-stone" valorization of waste into a customized catalyst in order to boost the parent biohydrogen process. For deep insights and detailed background of the FRs-based BCs properties, the full characterization details of SEM, FTIR, XRD, and BET/BJH are provided in the Supplementary Information.

### 3.9 Influence of FRs-based biochar on PFHP matrices

The effect of the optimized doses of all FRs-based biochar on cumulative hydrogen production and the hydrogen production rate is illustrated in Fig. 6(a) and Fig. 6(b), respectively. All FRs-based BCs samples were tested with a view to evaluating the optimal concentration range from 0 mg/L to 100 mg/L, at intervals of 20 mg/L between concentrations. The divergent behavior displayed by all the differing BCs at different optimal concentrations (Table 9) was due to their unique catalytic behaviors. Optimal dose of 80 mg/L, 40 mg/L, 60 mg/L, and 60 mg/L was observed for BC1, BC2, BC3 and BC4 respectively.

While examining the PFHP yield, the control reactor (CG; 0 mg/L) achieved a CHP of 158.07 mL; however, incorporation of all BCs was vastly superior to this control, with BC1 achieving a yield of 380 mL and BC2 achieving 530 mL. In comparison, BC3 attained the maximum output of the group yielding 570 mL, most likely due to its ultra-microporous structure (185 $m^2$/g surface area, 3.8 nm pore diameter) and well-balanced composition of minerals (8.2% CaO, 4.5% MgO), which aided in promoting the bacteria activity and provided a buffering capacity for the pH during stationary metabolic stages [104, 105]. BC4, though achieving 397.6 mL, exploiting its graphitic domains (high peaks in its XRD pattern at 26.4° and 42.8°) in facilitating electron shuttling, was diminished in yield due to a lower accessibility of surface area as compared with BC3. The superiority shown by BC3, slightly better than BC2 by 7.5% and BC1 by 50% is due to a synergistic effect of the oxygenated functional groups (FTIR peaks: C=O in the region of 1711 $cm^{-1}$) and the inorganic composition acting together to reduce metabolic inhibition [102, 106].

Concerning HPR, all forms of BCs yield a higher catalytic power than that of CG, as the control reactor had a production rate of 6.96 mL/h, which was maintained for a period of 36 hours during maximum metabolic growth. However, in contrast, the BCs incorporation produced a marked acceleration in kinetics, with BC1 exhibiting the highest HPR of 14.5 mL/h for a period of 48 hours. Whereas BC2 yielded 11.5 mL/h (over a period of 36 hours), its vesicular macropores (20.15 nm) limited microbial activity enhancement. In contrast, BC3 peaked at 13.5 mL/h during 48 hours due to its single-microporous surface that supported sustained microbial activity [102, 106]. Moreover, the BC4 achieved the highest rate of 14.91 mL/h over a period of 36 hours, mainly attributed to its graphitic carbon crystal lattice and its orderly aligned mesoporous lattice (4.31 nm), which served to promote the direct interspecies electron transfer mechanism manifested in the effect of the hydrogenase-type enzymes [107]. In all biochar reactions exhibited a rate greater than CG which represented a 57% boost over the BC1 yield and a 30% boost over the BC2 yield, clearly highlighting the vital importance which is played by the aspects of crystalline order of BCs (As evidenced via XRD; an unusually sharp peak appears at 44.5°) serving to promote rapid conduction of electronic flow to the site of enzymatic activity. These findings closely match previous studies using biomass-based biochar in biohydrogen production systems. The outstanding physicochemical properties of FRs-based biochar, including high surface area, contaminant immobilization, microbial growth support, nutrient availability, buffering capacity, and enhanced electron transfer, make it an excellent catalyst for biohydrogen production. Also, these FRs-based biochar produced a higher yield in the PFHP system due to its greater catalytic efficiency compared to standard biomass-based biochar [21, 24-27].

### 3.10 Influence of all BC on the microenvironment of fermentation broth

Reducing sugar (RS) is a building block in PFHP as microbes consume RS and convert it into VFAs and subsequently into biohydrogen, so a higher RS yield result in higher PFHP efficiency. Experimental values of RS with optimized doses of all biochar and CG are presented in Fig. 6(c), which shows that the CG accumulated the lowest concentration, 2.26 g/L of RS, but biochar differentially modulated hydrolysis, which might be through pore-driven enzyme immobilization and surface chemistry effects. A highest RS concentration of 2.65 g/L, 2.75 g/L, 2.7 g/L, and 3.19 g/L was observed in BC1, BC2, BC3 and BC4 respectively. All four biochar types outperformed CG as reactors with biochar amendment gave 17% to 41% higher RS yield than CG. From all biochar types, BC4 shows extraordinary performance over its counterparts and gave 20%, 16% and 18% RS yield in comparison to BC1, BC2, and BC3, respectively. These results indicated that BC4 shows higher enzymatic efficiency and higher superiority for the growth of hydrolytic bacteria because a higher yield of RS is directly attributed to hydrolytic bacteria [108]. Along with RS yield, residual concentration of RS at the end of the PFHP is also an important parameter correlating with the acidogenic bacterial performance, as the higher the residual RS, the lower the VFAs accumulation, meaning lower acidogenic bacterial performance [109]. When analyzing residual RS, CG shows a maximum residual RS of 0.62 g/L, while BC1, BC2, BC3, and BC4 resulted in 0.45 g/L, 0.56 g/L, 0.47 g/L, and 0.55 g/L, respectively, demonstrating superior sugar consumption by BC1, achieving 18% and 20% lower RS residual concentration than BC4 and BC3, respectively. BC1 extraordinary RS consumption might be due to enhanced cellulase binding on its macropore-dominated surface (36.24 nm pores, 6.14 m$^2$/g) along with exposed mineral phases (SEM: quartz/calcite aggregates), which provided nucleation sites for enzymatic activity [110]. BC3 microporous framework (3.80 nm) balanced these effects (0.47 mg/L residual sugar) by concentrating hydrolytic enzymes for enhancing microbial activity, while its oxygen functionalities (FTIR: 1711 cm$^{-1}$ C=O) might minimize the non-productive enzyme adsorption [108, 110].

The concentration of VFAs (acetic acid (AA), butyric acid (BA), propionic acid (PA), and bioethanol (BE)) in CG and reactors amended with an optimized dose of all four types of biochar are presented in Fig. 6(d, e, f, g). During 12–24 h, the intermediate by-product concentration increased rapidly owing to the conversion of sugar to hydrogen, acetic acid, butyric acid, and other substances (Eqs. (25) – (30)). In the CG, the levels of butyric acid and acetic acid exhibit a decrease from 24 h to 72 h. In contrast, a sharp reduction in the presence of BC1, BC2, BC3, and BC4 was observed in 24-36 h, suggesting a faster and efficient conversion of produced VFAs into biohydrogen. A sum of AA and BA can be used to represent the overall efficiency of the PFHP system, as AA and BA are the two most feasible metabolic pathways to produce a higher amount of biohydrogen [111, 112]. At the same time, PA and BE-based pathways yield lower biohydrogen in the PFHP system due to the stoichiometry of the systems [111, 113, 114]. However, adaptation of metabolic pathways is very complicated, as smaller changes in feedstock, pH, ORP, reactant/product concentration, catalyst, and other factors can heavily influence it and migrate to different types of metabolic pathways [115]. Therefore, based on the produced metabolite concentrations, in this work, it was observed that the PFHP system adopted 3 different types of fermentation pathways, which could be classified into propionate-type (AA+PA), butyrate-type (AA+BA) [116], and ethanol-type fermentation (AA+BE) [117].

$$C_6H_{12}O_6 + 2H_2O \rightarrow 2C_2H_4O_2 + 2CO_2 + 4H_2 \tag{5}$$

$$C_6H_{12}O_6 + 2H_2 \rightarrow 2C_3H_6O_2 + 2H_2O \tag{6}$$

$$C_6H_{12}O_6 \rightarrow C_4H_8O_2 + 2CO_2 + 2H_2 \tag{7}$$

$$C_2H_4O_2 + 2H_2O \rightarrow 4H_2 + 2CO_2 \tag{8}$$

$$C_4H_8O_2 + 6H_2O \rightarrow 10H_2 + 4CO_2 \tag{9}$$

$$C_6H_{12}O_6 \rightarrow C_2H_6O + 2CO_2 \tag{10}$$

Overall, all biochar gave higher AA+BA results when compared to (CG 0.87 g/L), the highest values of 0.9 g/L, 1.17 g/L, 1.32 g/L, and 2.36 g/L were observed for BCI, BC2, BC3, and BC4, respectively, which were 3.45%, 34.48%, 51.72%, and 171.26% higher than CG. These results suggested that biochar boosted acetic acid and butyric acid production according to the different intrinsic properties of biochar; the previous studies strengthen these findings [54, 112]. Similarly, all biochar gave higher AA+PA results when compared to (CG 1.04 g/L), the highest values of 1.05 g/L, 1.17 g/L, 1.4 g/L, and 2.41 g/L were observed for BCI, BC2, BC3, and BC4, respectively, which were 0.96%, 12.50%, 34.62%, and 131.73% higher than CG. Surprisingly, except BC4 all biochar gave lower AA+BE results when compared to (CG 0.81 g/L), the highest values of 0.4 g/L, 0.5 g/L, 0.55 g/L, and 2.11 g/L were observed for BCI, BC2, BC3 and BC4, respectively, which were 50.62%, 38.27% and 32.10% lower than CG, only BC4 show 160.49% higher value than CG. Due to the unique intrinsic characteristics of all four biochar to influence pH, ORP, and other parameters in PFHP system, they adopted different metabolic pathways, all kind of biochar did not follow the ethanol type fermentation because it normally occur at pH < 4.5 which was not observed, also all reactor gave lower results for AA+BE along with previous study also strengthening our claims [115]. When analyzing AA+BA and AA+PA, considerable variations were observed as BC1 and BC3 followed the butyrate type fermentation, while BC2 and BC4 followed the propionate type fermentation. This change in metabolic pathway can also be justified based on PFHP yield, as even though BC4 produced a higher amount of VFAs than BC3, by following the stoichiometrically lower yield metabolic pathway, the overall yield of the BC4-amended system decreased from BC3.

**Table 9:** Fundamental characteristics of FRs-based biochar and their influence on PFHP parameters at optimal dose.

| Key Biochar Properties Influencing PFHP Performance | | | | |
|---|---|---|---|---|
| Biochar | Precursor | BET ($m^2/g$) | Pore Size (nm) | Key Characteristics |
| BC1 | S1 | 7.20 | 6.74 | High ash (27.57%), macropores |
| BC2 | S2 | 5.98 | 20.15 | High volatiles (65.85%) |
| BC3 | S3 | 185.14 | 4.67 | Highest microporosity, O-rich (O/C=1.77) |
| BC4 | S4 | 76.58 | 4.31 | Mesoporous, graphitic domains |
| PFHP Performance at Optimal Doses | | | | |
| Parameter | BC1 | BC2 | BC3 | BC4 |
| Optimal Dose | 80 mg | 40 mg | 60 mg | 60 mg |
| Peak HPR (mL/h) | 9.5 | 11.5 | 13.5 | 14.9 |
| Time to Peak | 36h | 36h | 48h | 36h |
| Total CHP (72h) | 380 mL | 530 mL | 570 mL | 398 mL |
| $H_2$/Total Gas | 25% | 30% | 37% | 33% |
| pH Stability | 5.9→6.1 | 5.9→6.1 | 5.75→6.0 | 6.06→6.19 |
| ORP (mV) | -215→-230 | -280→-360 | -300→-390 | -261→-312 |
| **Metabolic Byproducts & Substrate Utilization (At peak HPR timepoints)** | | | | |
| **Fermentation Broth Component** | **BC1** | **BC2** | **BC3** | **BC4** |
| Acetic Acid | 0.65 mg/L | 0.85 mg/L | 0.85 mg/L | 0.38 mg/L |
| Butyric Acid | 0.18 mg/L | 0.38 mg/L | 0.45 mg/L | 0.62 mg/L |
| Propionic Acid | 1.15 mg/L | 1.25 mg/L | 1.20 mg/L | 0.45 mg/L |
| Bioethanol | 0.90 mg/L | 1.00 mg/L | 0.92 mg/L | 0.77 mg/L |
| RS Consumption | Slow | Moderate | Rapid | Moderate |

When analyzing pH trends of the current study, the CG achieved a lowest pH of 5.82 within 24 hours, while biochar-amended reactors depicted a higher drop in pH, as depicted in Fig. 6(h), but this drop was much smoother due to the distinct buffering mechanisms [27, 118]. The reactor amended with BC1 achieved a pH of 5.50 at 36 hours, while BC2 stabilized at pH 5.70, limited by its volatile-driven pore collapse (5.98 $m^2$/g surface area); meanwhile, BC4 reached pH 5.51, utilizing its graphitic layers for partial metabolite adsorption. The BC3 outperforming CG and all other types FRs-based biochar, exhibited the greatest buffering efficiency in combating acidic wastes developed by the biological activity, effecting a pH retention of 5.45 via virtue of its great carbonate equivalence (28.3 mg $CaCO_3$/g) as well as its highly alkaline metal oxides (CaO/MgO) which sustained the general scenario of alkalinity which effectively continues to preserve the optimum pH range of hydrogenase enzymatic activity of 5.2-5.8 [119]. This performance was superior to that of BC1 by 7% and BC2 by a mere 0.9%. It was closely observed with all results being immediately justifiable due to the cross-linking catalyzed efficiency via oxygen-containing functionalities that are inherently prominent in the BC3 (O/C = 1.77), as well as the retention of the buffering ions in the domains of the micropores being deeply related. All of the experimental groups displayed a very similar phenomenon, explicating in detail that even CG displayed in fact the minimal possible pH values (between 5.45-6.0) producing clear relationships with relevant correlation to previous researches, such as some other studies who have noted that maximum production of biohydrogen gas mildly takes place within a slightly acidic media, between the acid-base pH range of 5.0 and some satisfactory degree to 6.5 [21, 24-27, 120, 121]. Hence, the acid conditions favor the metabolic networking, which is important in supporting the flow of hydrogen and volatile fatty acid production, such as butyrate or acetate, towards sustenance of the hydrogenotrophic pathway, with inhibition of competing pathways like propionate [121]. The drop in pH that happened during the acidogenic period of HPR noted above is directly related to the great increase in accumulation of volatile fatty acids, which acids are subsequently used for the improvement of biohydrogen gains, indicating that a lower pH is indeed contributive to higher yield of PFHP [120, 121]. Obviously, therefore, the modification of the ORP levels of the broths involves important results concerning the metabolic activities of bacteria as well as their interaction with means of both bacterial metabolisms being crucial ingredients in biohydrogen production [122]. One can note that the hydrogen yields of biochar were all competitive; however, as was pointed out, therefore, the biochar yield HPR was greater.

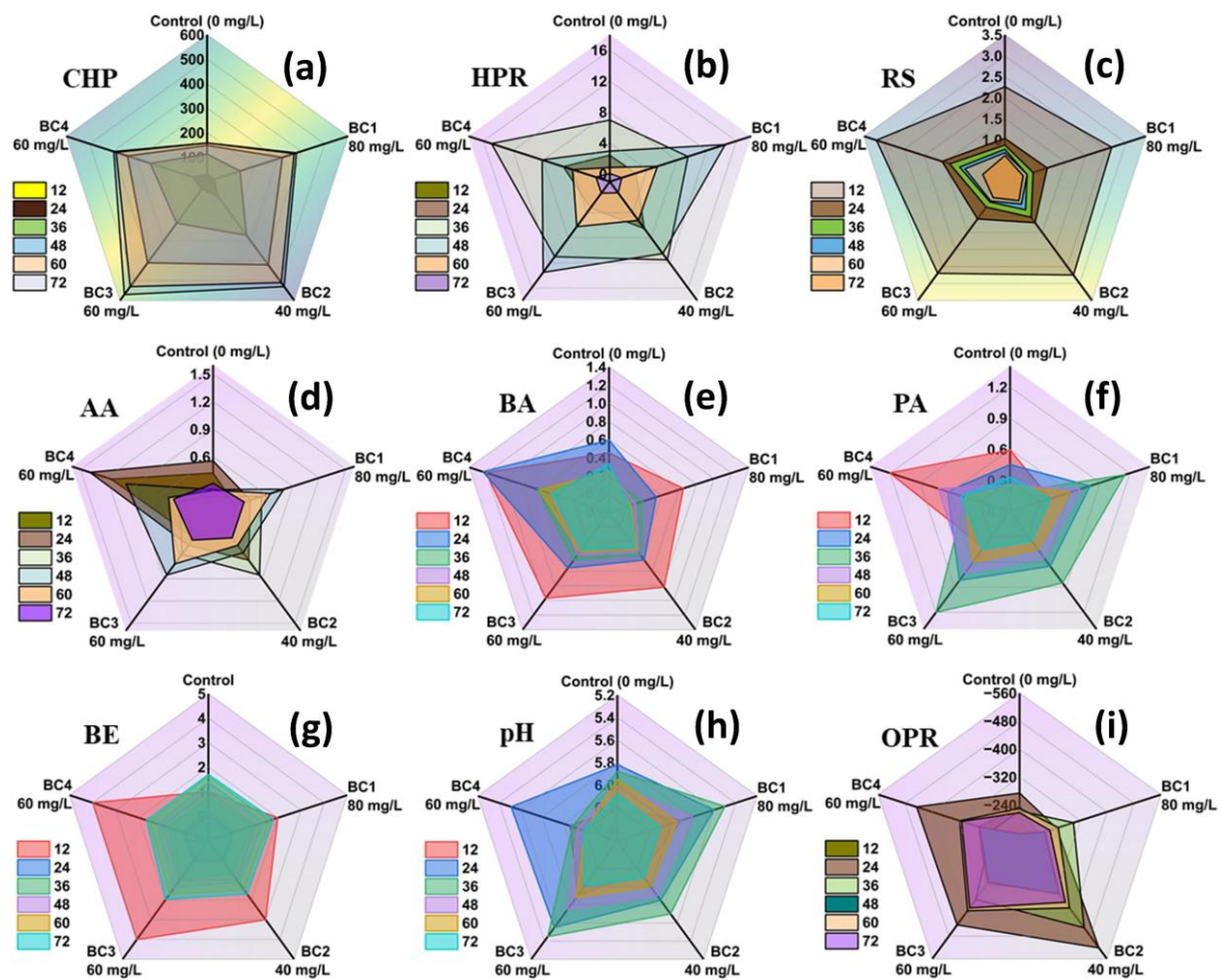

**Fig. 6:** Influence of all BC on PFHP parameters at optimized doses (a) CHP, (b) HPR, (c) RS, (d) acetic acid, (e) butyric acid, (f) propionic acid, (g) bioethanol, (h) pH, (i) ORP.

During PFHP, a higher drop in biochar amended reactors was observed in comparison to the CG, which directly correlates with efficient ORP management to enhance PFHP yield using all types of biochar over CG, as shown in Fig. 6(i). Because from the start of the fermentations until the beginning of the biohydrogen production, a drop in the ORP is observed, attributed to the activity of hydrolytic bacteria [108]. The CG stabilized at an ORP of -274 mV by 24 hours, while biochar-augmented reactors achieved substantially lower potentials due to their distinct redox-active properties [27]. Generally, the ORP value close to -240 mV favors the development of facultative anaerobic bacteria, while lower ORPs favor the development of strict anaerobic bacteria [123]. The decrease in ORP values was greatest with BC2 (-521 mV), which exceeded BC3 values by 16% and BC4 by 74%. This was due mainly to the peaks in FTIR of phenolics (FTIR: 1575 cm-1 C=C/C=N) as they scavenged the dissolved $O_2$ extremely rapidly (< 12 h post inoculation). The extreme efficiency of reduction was due to the vesicular pore structure (SEM: 20 nm spherical voids) and specific surface area (5.98 $m^2$/g), permitting microbial growth. At the same time, its lignin-derived redox functionalities accelerated oxygen consumption. In contrast, BC4's graphitic domains (XRD: 26.4° peak) achieved moderate reduction (-446 mV) through electron shuttling, whereas BC3's microporous structure (185.14 m²/g) prioritized pH buffering over ORP control despite its oxygen-rich matrix (O/C=1.77) [118]. These lower ORP values achieved by all biochar-amended reactors are characteristic of the potentials given in literature for hydrogenases [123-125]. During the final hours of the fermentation, an increase in ORP values was noted, probably indicating bacterial death [108]. This study successfully establishes a closed-loop, zero-waste biorefinery model by demonstrating the synergistic integration of photo-fermentative biohydrogen production (PFHP) with the thermochemical valorization of its fermentation residues (FRs). The core innovation lies in transforming lignocellulosic waste into engineered biochar (BC) catalysts that are subsequently recycled to enhance the initial biological process, thereby achieving a circular economy approach for sustainable energy production [5, 10].

**Conclusion**

This work successfully demonstrates a circular biorefinery model whereby the residues from the photo fermentation phase during the biohydrogen production cycle are converted to the engineered biochar, which can then transform the output of the first process to greatly improve its evident efficiency. The pyrolysis of the product at 700°C effectively tailored the biochar properties of the product: it was shown that microporous BC3 was of advantage for total H$_2$ yield enhancement through stimulation of the stable pH conditions, whilst graphitic BC4 greatly enhanced production rates before reliance on electron transfer. Kinetic and thermodynamic studies clearly confirmed the diffusion-dominated mechanisms and the energy aspects of the net gain of improvements in properties of materials through the use of the precursors involved. The reintroduction of biochar thus effected positive improvements in H$_2$ metrics, but moreover allowed the fine-tuning of the metabolic pathways through regulation of products and their fermentation conditions as discussed above. The effectively circular route can thus be said to combine the waste product nature of the materials to produce very high value products (syngas, bio oil, biochar catalysts) and thus provide a route to the near 100% recovery of all the solid phase resources from the system. The work has given a very strong validation of the possibilities of the FRs-based biochar as a multi-purpose catalyst for use in sustainable bioenergy systems for environmentally and economically viable waste reduction operation towards the even more sustainably integrated thermochemical and biological methods if valorization is to be successfully achieved in a zero-waste biorefinery.